\documentclass[
    reprint,
    amsmath,amssymb,
    aps,
]{revtex4-2}

\usepackage{array}[=2016-10-06]

\usepackage{graphicx}
\usepackage{subcaption}
\usepackage{dcolumn}
\usepackage{bm}
\usepackage{orcidlink}
\usepackage{caption}
\usepackage{booktabs}
\usepackage{multirow}
\usepackage{tabularx}
\usepackage{url}
\usepackage{verbatim}
\usepackage{tikz}
\usetikzlibrary{shapes.geometric, arrows.meta, positioning}
\usepackage{ragged2e}

\begin{document}

\preprint{APS/123-QED}

\title{Threshold quantum state tomography of two-qubit polarization-entangled photons generated via type-0 SPDC}

\author{Vikas Tanwar\,\orcidlink{0009-0003-3168-5794}}
\affiliation{School of Quantum Technology, Defence Institute of Advanced Technology, Pune, India.}%Lines break automatically or can be forced with \\
\author{Gayatri Thik\,\orcidlink{0009-0005-0360-8032}}
\affiliation{School of Quantum Technology, Defence Institute of Advanced Technology, Pune, India.}

\author{Makham Chakradhar\,\orcidlink{0009-0008-0454-0596}}
\affiliation{School of Quantum Technology, Defence Institute of Advanced Technology, Pune, India.}

\author{Satya Bhaskara Pramod Karasi\,\orcidlink{0009-0001-8312-3908}}
\affiliation{School of Quantum Technology, Defence Institute of Advanced Technology, Pune, India.}

\author{Raghavan G\,\orcidlink{0009-0008-8311-7840}}
\affiliation{School of Quantum Technology, Defence Institute of Advanced Technology, Pune, India.}

\author{Srinivasan K\,\orcidlink{0000-0002-7707-7896}}
\email{Corresponding Author: srinik@diat.ac.in}
\affiliation{School of Quantum Technology, Defence Institute of Advanced Technology, Pune, India.}

\date{\today}% It is always \today, today,
             %  but any date may be explicitly specified

\begin{abstract}

Quantum state tomography (QST) provides a complete characterization of experimentally generated quantum states. However, the number of measurements it entails, increases exponentially with system size. Threshold quantum state tomography (tQST) offers a measurement-efficient alternative by selectively targeting density-matrix elements that are expected to contribute most significantly to the reconstructed state, guided by positivity constraints on the density matrix. We present an experimental implementation of tQST for polarization-entangled photon pairs generated via type-0 spontaneous parametric down-conversion in a periodically poled potassium titanyl phosphate (PPKTP) crystal in a Sagnac interferometer. We reconstruct all four maximally entangled Bell states using only six projective measurements, compared with the sixteen projectors required for conventional two-qubit QST, corresponding to a $62.5\%$ reduction in the number of measurements. The six-projector tQST reconstruction yields fidelities of $0.9235$ and $0.9586$ for $|\Phi^{+}\rangle$ and $|\Psi^{+}\rangle$, respectively, while fidelities of $0.9792$ and $0.9847$ are obtained for $|\Phi^{-}\rangle$ and $|\Psi^{-}\rangle$, respectively. We further find that fidelities above $0.98$ can be achieved for all four Bell states either by augmenting two additional projective measurements or by employing a state-estimation algorithm based on the sequential least-squares programming (SLSQP). These results demonstrate the potential of tQST for resource-efficient characterization of polarization-entangled photonic states while highlighting the trade-off among measurement reduction, reconstruction accuracy, and statistical precision.

\end{abstract}

%\keywords{Suggested keywords}%Use showkeys class option if keyword
                              %display desired
\maketitle

%\tableofcontents

\section{\label{sec:level1}Introduction}%\protect\\ The line
%break was forced \lowercase{via} \textbackslash\textbackslash}

Accurate characterization of experimentally generated quantum states is essential for the development and validation of quantum technologies, including quantum communication, quantum computing, and quantum-enhanced sensing~\cite{10.1117/12.504402}. Quantum state tomography (QST) is the standard and most widely used technique for obtaining a complete description of an unknown quantum state ~\cite{PhysRevA.64.052312}. In QST, the density matrix of a quantum system is reconstructed from a series of measurements performed on an ensemble of identically prepared copies of the state.  However, the measurement resources required for conventional QST increase exponentially with the number of qubits, making complete state reconstruction increasingly challenging for large quantum systems. In the measurement convention considered here, a complete tomography of an $n$-qubit system requires $4^n$ projective measurements ~\cite{Altepeter2004,PhysRevA.64.052312}.

Several measurement-efficient approaches have therefore been developed to reduce the experimental resources required for quantum-state reconstruction. These include compressed-sensing-based tomography~\cite{PhysRevLett.105.150401}, Bayesian tomography~\cite{Lu2022Bayesian,BUZEK1998454}, adaptive tomography~\cite{Qi2017}, and other methods that exploit prior knowledge about the quantum state or target specific properties of interest. Although such approaches can substantially reduce the measurement requirements, they may rely on assumptions about the state, its rank, or the particular property being estimated. Consequently, developing a measurement-efficient method that reconstructs the complete density matrix without prior knowledge of the quantum state remains an important challenge.

To address this problem, the threshold quantum state tomography (tQST) framework was introduced. This method exploits the positive semidefinite nature of the density matrix, $\hat{\rho}\geq 0$, which imposes the constraint
\begin{equation}
|\rho_{ij}|\leq\sqrt{\rho_{ii}\rho_{jj}}
\end{equation}
The diagonal elements of the density matrix are first determined, and the above relation is then used to obtain an upper bound on the magnitude of each off-diagonal element. For a chosen threshold $t$, only those off-diagonal elements whose upper bounds satisfy $\sqrt{\rho_{ii}\rho_{jj}}\geq t$ are considered potentially significant and are therefore selected for further measurements. The projective measurements associated with the remaining elements are ignored. Thus, unlike conventional QST, in which a fixed tomographically complete set of measurements is performed, the number of measurements in tQST is adapted to the experimentally generated quantum state. In the limit $t\rightarrow0$, tQST approaches conventional QST because all required projective measurements are retained. The measurement reduction is expected to be particularly significant for states whose density matrices contain many zero or small-valued elements. This tQST  was first introduced by Binosi et al.~\cite{10.1063/5.0219143},  who demonstrated its experimental feasibility on the seven-qubit IBM Quantum processor by reconstructing states generated by random quantum circuits. They also showed numerically, that a seven-qubit $W$ state can be reconstructed using only 170 projective measurements, compared with 16,384 measurements for conventional QST. Subsequently, Garberoglio et al.~\cite{PhysRevA.111.032436} introduced enhanced compressive threshold quantum state tomography (ECT-QST), which combines thresholding with compressed-sensing-inspired optimization of measurement settings and extends the framework to arbitrary qudit systems. They demonstrated the reconstruction of GHZ, $W$, and random states for systems containing up to seven qubits, with the required number of optimized measurement settings exhibiting favorable scaling with system size ~\cite{PhysRevA.111.032436, Greenberger1989, PhysRevLett.94.070402, PhysRevA.62.062314}. More recently, Caruccio et al.~\cite{Caruccio2025} experimentally validated tQST on a fully reconfigurable integrated photonic platform using up to four photonic qubits encoded in spatial modes. Their experiments demonstrated a substantial reduction in the number of projective measurements while maintaining high-fidelity state reconstruction.
provides a measurement-efficient approach for reconstructing the density matrix without requiring prior knowledge of the state ~\cite{10.1063/5.0219143, Caruccio2025}. 

Spontaneous parametric down-conversion (SPDC) is one of the most widely used techniques for generating polarization-entangled photon pairs and multi-photon entangled states, which constitute important resources for quantum communication and quantum information processing.~\cite{PhysRevA.60.R773, PhysRevLett.75.4337, Sergienko01072003,Bourennane2003,10.1063/5.0023103, PhysRevLett.90.200403,PhysRevA.64.010102,Gaertner2003High}. In particular, type-0 phase-matched periodically poled potassium titanyl phosphate (PPKTP) crystals incorporated in Sagnac interferometric geometries provide bright and stable sources of polarization-entangled photon pairs~\cite{y1pg-nx49}. Complete characterization of a two-qubit polarization state using conventional QST requires 16 projective measurements in the measurement convention adopted here. However, despite the recent demonstrations of tQST on superconducting and integrated-photonic~\cite{Caruccio2025,10.1063/5.0219143}, to the best of our knowledge, its experimental implementation for two-qubit polarization-entangled photons generated through SPDC has not been reported. Bright photon sources have the advantage that they facilitate reliable tQST with high fidelity and visibility of all four Bell states. 

In this work, we experimentally implement threshold quantum state tomography (tQST) for polarization-entangled photon pairs generated through SPDC in a Sagnac interferometer incorporating a type-0 periodically poled potassium titanyl phosphate (PPKTP) crystal \cite{Thik_2026}. We demonstrate the reconstruction of the density matrices of all four Bell states using only six projective measurements, compared with the 16 measurements employed in conventional standard quantum state tomography (sQST). The reconstructed Bell states exhibit fidelities exceeding $0.9235$ with their corresponding ideal states. Thus, the six-projector tQST protocol achieves a $62.5\%$ reduction in the number of projective measurements while maintaining high reconstruction fidelities. To further improve the reconstruction fidelities, particularly for the $|\phi^+\rangle$ and $|\psi^+\rangle$ states, we investigate two approaches: (i) extending the tQST protocol from six to eight projective measurements and (ii) employing a density-matrix reconstruction procedure based on sequential least-squares programming (SLSQP). We systematically evaluate the fidelity, purity, and concurrence of the reconstructed states and compare the performance of the different tQST configurations with conventional sQST. Our results demonstrate the applicability of measurement-efficient tQST to polarization-entangled photon sources and highlight its potential for reducing the experimental resources required for quantum-state characterization in photonic quantum information systems.

The remainder of this paper is organized as follows. Section II presents the theoretical framework of threshold quantum state tomography and the density-matrix reconstruction procedures. Section III describes the experimental setup for generating all four polarization-entangled Bell states through type-0 SPDC in a Sagnac interferometer. Section IV presents the experimental results and a comparative analysis of the reconstructed quantum states obtained using conventional sQST and the different tQST approaches. Finally, Sec. V summarizes the main findings and conclusions.

\section{\label{sec}Theoretical Background of Threshold Quantum State Tomography}
In this section, we summarize the theoretical framework of threshold quantum state tomography (tQST) with particular emphasis on its application to two-qubit quantum systems~\cite{10.1063/5.0219143}. We describe how the positivity constraints on the density matrix are exploited to identify the relevant off-diagonal elements and select the projective measurements required for quantum-state reconstruction.

The primary objective of quantum state tomography (QST) is to reconstruct the density matrix of an experimentally prepared quantum system \cite{Altepeter2004}. This is achieved by performing a series of projective measurements on an ensemble of identically prepared copies of the quantum state. For a physical quantum state, the density matrix $\hat{\rho}$ must satisfy three fundamental conditions: it must be Hermitian, have unit trace, and be positive semidefinite~\cite{nielsen2010quantum,Altepeter2004}, i.e.,
\begin{equation}
\hat{\rho}=\hat{\rho}^{\dagger}, \qquad
\operatorname{Tr}(\hat{\rho})=1, \qquad
\hat{\rho} \geq 0.
\end{equation}
The quantity $\operatorname{Tr}(\hat{\rho}^{2})$ represents the purity of the state and satisfies
\begin{equation}
\frac{1}{d}\leq \operatorname{Tr}(\hat{\rho}^{2})\leq1
\end{equation}
for a $d$-dimensional quantum system. In practice, experimental imperfections and statistical fluctuations can cause a directly reconstructed density matrix to deviate from the physical-state constraints. Reconstruction methods such as maximum-likelihood estimation~\cite{PhysRevA.68.012305,PhysRevA.61.010304} can therefore be employed to obtain a physical density matrix. In the conventional sQST process complete reconstruction of an $n$-qubit system requires $4^n$ projective measurements~\cite{PhysRevA.64.052312, Altepeter2004}. Thus, the measurement resources required for conventional QST increase exponentially with the number of qubits, making complete characterization increasingly demanding for multiqubit quantum systems.

Threshold quantum state tomography (tQST) is particularly advantageous for quantum states whose density matrices are sparse or contain many small-valued elements. The basic principle of tQST follows from the positive semidefinite property of the density matrix. To illustrate this principle, consider a $2\times2$
principal submatrix $\mathbf{M}_{ij}$ of an arbitrary density matrix $\hat{\rho}$, corresponding to the
indices $i$ and $j$:
\begin{equation}
\mathbf{M}_{ij} = 
\begin{pmatrix}
\rho_{ii} & \rho_{ij} \\
\rho_{ji} & \rho_{jj}
\end{pmatrix}
\end{equation}
Since the density matrix $\hat{\rho}$ must be
positive semidefinite, its principal submatrix
$\mathbf{M}_{ij}$ must also be positive
semidefinite. Therefore, its eigenvalues must be
nonnegative, i.e., $\lambda_1,\lambda_2\geq 0$. The determinant of the matrix is related to its eigenvalues as,
\begin{equation}
    \det(\mathbf{M}_{ij}) = \rho_{ii}\rho_{jj} - \rho_{ij}\rho_{ji} = \lambda_1 \lambda_2 \geq 0.
\end{equation}
Using the Hermiticity condition $\rho_{ji}=\rho_{ij}^{*}$, we obtain
\begin{equation}
|\rho_{ij}|\leq\sqrt{\rho_{ii}\rho_{jj}}.
\label{eq}
\end{equation}

Equation~(\ref{eq}) is the fundamental relation exploited in tQST. It provides an upper bound on the magnitude of each off-diagonal element in terms of the corresponding diagonal populations. In particular, if a diagonal element $\rho_{ii}=0$, then Eq.~(\ref{eq}) implies that $\rho_{ij}=0$ for every $j$. Consequently, the entire $i$th row and column of the density matrix must vanish. Similarly, if $\rho_{ii}$ and $\rho_{jj}$ are nonzero but small, then the corresponding off-diagonal element $\rho_{ij}$ is necessarily bounded by the small quantity denoted by $B_{ij}=\sqrt{\rho_{ii}\rho_{jj}}$. Thus, the diagonal elements indicate which off-diagonal elements may have significant magnitudes. These significant off-diagonal elements are identified using a threshold determined from the diagonal entries. This property forms the basis for reducing the number of measurements required for density-matrix reconstruction.

 The Gini index (GI)~\cite{5238742} is used to determine the threshold $t$. Let $\mathbf{c}$ denote the vector formed by the diagonal elements of the density matrix,

\begin{equation}
    \mathbf{c}=(c_1,c_2,\ldots,c_N)
=(\rho_{11},\rho_{22},\ldots,\rho_{NN})
\end{equation}

where $N=2^n$ for an $n$-qubit system and $c_i\geq0$ for all $i$. The elements of $\mathbf{c}$ are arranged in ascending order and denoted by

\begin{equation}
    c_{(1)}\leq c_{(2)}\leq\cdots\leq c_{(N)}
\end{equation}

The Gini index is then defined as
\begin{equation}
\text{GI}(c) = 1 - 2 \sum_{k=1}^N \frac{c_k}{\|c\|_1} \left( \frac{N - k + \frac{1}{2}}{N} \right)
\label{eq:Gini1}
\end{equation}
where

$$
\|\mathbf{c}\|_1=\sum_{i=1}^{N}c_i
$$

is the $L_1$ norm of $\mathbf{c}$. The Gini index satisfies

$$
0\leq\mathrm{GI}(\mathbf{c})\leq1-\frac{1}{N}
$$

where the lower bound corresponds to the case in which all elements of $\mathbf{c}$ are equal, while values approaching the upper bound indicate increasing inequality among the elements.

For tQST, the vector $\mathbf{c}$ is therefore obtained directly from the experimentally measured diagonal populations of the density matrix. The threshold is chosen according to
\begin{equation}
t=
|\mathbf{c}|_1
\frac{\mathrm{GI}(\mathbf{c})}{2^n-1}
\label{Eq:thresholdE1}
\end{equation}
 Since the components $c_i=\rho_{ii}$ are nonnegative and the density matrix has unit trace, we obtain $\|\mathbf{c}\|_1=\sum_{i=1}^{N}\rho_{ii}=1$ and Eq.~(\ref{Eq:thresholdE1}) can equivalently be written as

\begin{equation}
  t=\frac{\mathrm{GI}(\mathbf{c})}{2^n-1}
\label{Eq:thresholdE2}  
\end{equation}

Therefore, the threshold is determined directly from the distribution of the experimentally measured diagonal populations. A greater inequality among these populations yields a higher Gini index and, consequently, a higher threshold. Once the diagonal elements have been measured in the computational basis, the threshold $t$ is calculated from Eq.~(\ref{Eq:thresholdE2}), and the off-diagonal elements satisfying $\sqrt{\rho_{ii}\rho_{jj}}\geq t$ are selected for further characterization. The corresponding projective measurements are then performed only for the selected elements. In the following subsections, we describe the procedure for identifying the projectors associated with these selected off-diagonal elements.

\begin{table}[ht]
\centering
\begin{tabular}{|c|c|c|c|}
\hline
$\hat{P}_{HH}$ & $\hat{P}_{HD}+i\hat{P}_{HR}$  & $\hat{P}_{DH}+i\hat{P}_{RH}$ & $\hat{P}_{DD}+i\hat{P}_{DR}$ \\ \hline
$\hat{P}_{HD}-i\hat{P}_{HR}$ & $\hat{P}_{HV}$ & $\hat{P}_{RR}+i\hat{P}_{RD}$ & $\hat{P}_{DV}+i\hat{P}_{RV}$ \\ \hline
$\hat{P}_{DH}-i\hat{P}_{RH}$ & $\hat{P}_{RR}-i\hat{P}_{RD}$ & $\hat{P}_{VH}$ & $\hat{P}_{VD}+i\hat{P}_{VR}$ \\ \hline
$\hat{P}_{DD}-i\hat{P}_{DR}$ & $\hat{P}_{DV}-i\hat{P}_{RV}$ & $\hat{P}_{VD}-i\hat{P}_{VR}$ & $\hat{P}_{VV}$ \\ \hline
\end{tabular}
\caption{\justifying Projector assignment for the reconstruction of the elements of a two-qubit density matrix, adapted from Ref.~\cite{10.1063/5.0219143}. The projector is defined as $\hat{P}_{ij}=|ij\rangle\langle ij|$, where $i,j\in\{H,V,D,R\}$. In the table, an entry of the form $\hat{P}_{ab}+i\hat{P}_{cd}$ indicates that $\hat{P}_{ab}$ is used to obtain information about the real part of the corresponding off-diagonal density-matrix element, while $\hat{P}_{cd}$ is used to obtain information about its imaginary part. For example, with the computational basis ordered as $\{|HH\rangle, |HV\rangle, |VH\rangle, |VV\rangle$\}, the matrix element $\rho_{12}$ is associated with the entry $\hat{P}_{HD}+i\hat{P}_{HR}$. Thus, $\hat{P}_{HD}$ is used for the reconstruction of $\operatorname{Re}(\rho_{12})$, while $\hat{P}_{HR}$ is used for the reconstruction of $\operatorname{Im}(\rho_{12})$.}  
\label{tab:Projectors_matrix}
\end{table}

\subsection{Projectors for off-diagonal entries of the density-matrix elements}
The real and the imaginary part of an off-diagonal element $\rho_{ij}$ of a density matrix $\hat{\rho}$ can be obtained from,
\begin{equation}
\rho_{ij}^{\text{re,im}} = \text{tr}(\Theta_{ij}^{\text{re,im}} \hat{\rho})
\end{equation}

where $\Theta_{ij}^{\mathrm{re}}$ is a $2^n\times2^n$ Hermitian operator with matrix elements equal to $1/2$ at the $(i,j)$ and $(j,i)$ positions and zero elsewhere. Similarly, $\Theta_{ij}^{\mathrm{im}}$ is a $2^n\times2^n$ Hermitian operator with matrix elements equal to $i/2$ at the $(i,j)$ position and $-i/2$ at the $(j, i)$ position, with all other matrix elements equal to zero. The experimentally accessible projectors associated with the real and imaginary parts of the density-matrix element $\rho_{ij}$ are then identified as follows:

\begin{equation}
P_{ij}^{\text{re,im}} = \arg \min_{\hat{P} \in \mathcal{P}} \|\Theta_{ij}^{\text{re,im}} - \hat{P}\|_F
\label{eq:proj_selection}
\end{equation}

where $\mathcal{P}$ denotes the set of all experimentally accessible projectors $\hat{P}$ for the two-qubit system, and $||\cdot||{F}$ represents the Frobenius norm~\cite{HornJohnson2012}. Using the procedure described above, the projector associated with the real or imaginary part of a selected off-diagonal element $\rho_{ij}$ is identified by minimizing the quantity in Eq.~(\ref{eq:proj_selection}) over all projectors $P\in\mathcal{P}$. The projector that yields the minimum Frobenius distance from the corresponding operator $\Theta_{ij}^{\mathrm{re}}$ or $\Theta_{ij}^{\mathrm{im}}$ is selected for the measurement of the associated density-matrix component. By applying this procedure only to the off-diagonal elements selected by the tQST threshold criterion, the number of projective measurements can be substantially reduced compared with conventional QST. The projectors associated with the selected density-matrix elements are listed in Table~\ref{tab:Projectors_matrix}.

The tQST protocol can be summarized by the following steps:
\begin{enumerate}
\item The diagonal elements $\rho_{ii}$ of the density matrix are first measured by performing projective measurements in the chosen computational basis.

\item A threshold $t$ is selected. Using the measured diagonal elements, the upper bound $\sqrt{\rho_{ii}\rho_{jj}}$ is calculated for each off-diagonal element. Only those off-diagonal elements satisfying
\begin{equation}
\sqrt{\rho_{ii}\rho_{jj}}\geq t
\end{equation}
are selected for further characterization.

\item A suitable set of projectors or observables associated with the selected off-diagonal elements is constructed, and only the corresponding measurements are performed.

\item Finally, the measurement results are used to reconstruct the density matrix, for example, through maximum-likelihood estimation.

\end{enumerate}

Unlike conventional QST, for which the measurement set is determined by the dimension of the Hilbert space, the number of measurements required by tQST depends on the selected threshold and the experimentally measured diagonal populations. In the limit $t\rightarrow0$, all off-diagonal elements are retained, and the protocol reduces to conventional QST. For $t>0$, off-diagonal elements whose upper bounds are smaller than the threshold are omitted, thereby reducing the number of measurements. Importantly, tQST does not require any a priori knowledge of the quantum state; instead, the measurements required after the initial diagonal measurements are selected using the information obtained from those measured populations~\cite{10.1063/5.0219143}. The threshold $t$ therefore controls the trade-off between the number of measurements and the amount of information retained in the reconstruction.

\section{Experimental Apparatus}

The polarization-entangled photon source employed
in this work is based on the experimental setup
developed and characterized in our previous
study~\cite{Thik_2026}.
In this section, we describe the experimental
arrangement for generating all four Bell states
using a single Sagnac-interferometer-based
entangled-photon source, along with the polarization
analysis and detection system employed for
quantum state tomography.

\begin{figure}[h]
\centering
\includegraphics[width=1\linewidth]{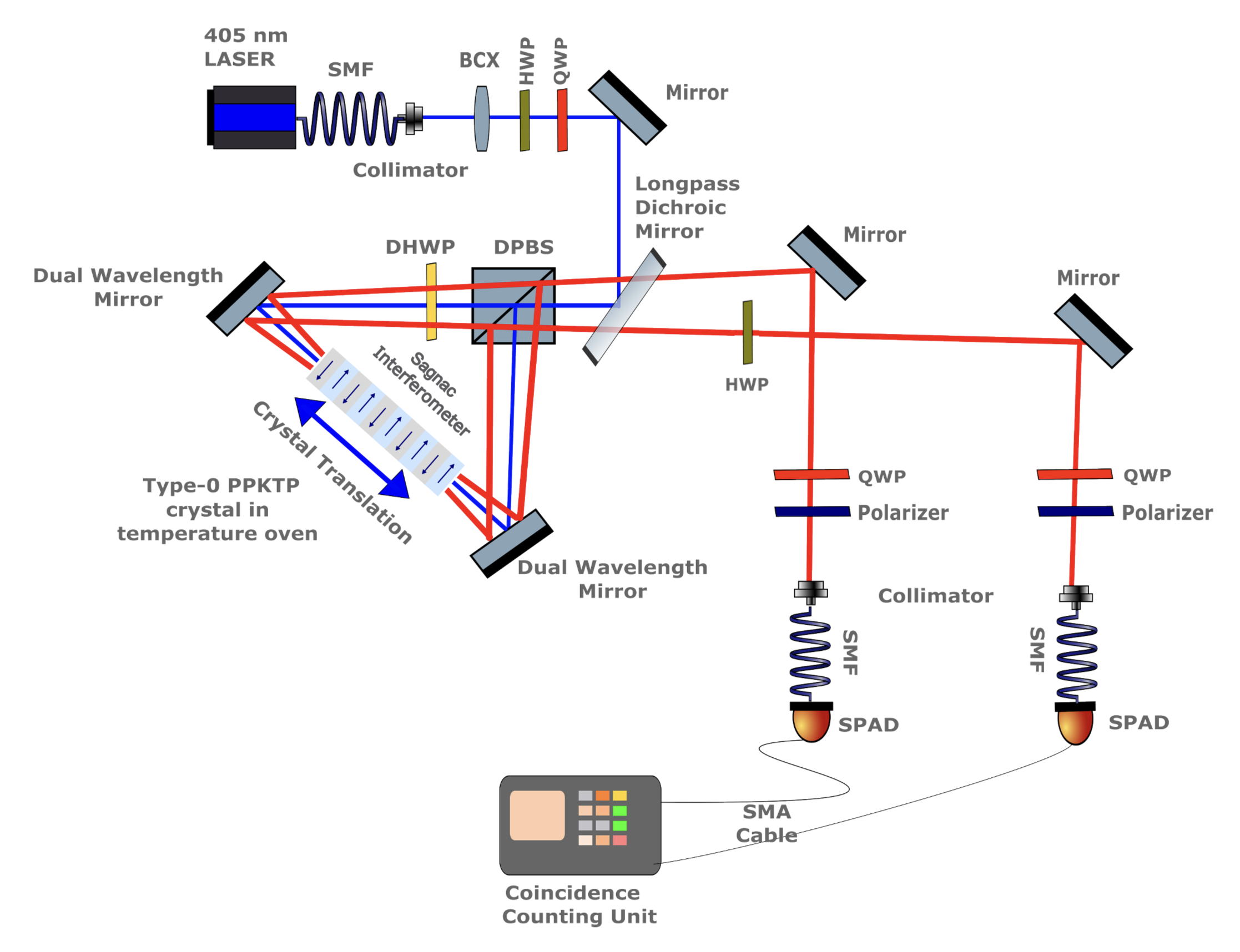}
\caption{\justifying Schematic of the experimental setup
for quantum state tomography of polarization-entangled
photon pairs. The entangled-photon source is
based on the type-0 PPKTP Sagnac interferometer
developed in our previous
work~\cite{Thik_2026}.
The signal and idler photons are analyzed using
a QWP and a polarizer in the signal and idler arms and detected using
single-photon avalanche diodes (SPADs).
}
\label{QST}
\end{figure} 

Polarization-entangled photon pairs are generated using a type - 0 phase-matched PPKTP crystal kept in the Sagnac interferometer as given in the schematic diagram in Fig. \ref{QST} ~\cite{BERA2025132401,Lee2015,Sundar:25,y1pg-nx49,Terashima2018,article,Li:15}. A diode laser with a central wavelength of $\lambda_p=405~nm$ and a spectral bandwidth of $0.018~nm$ is used as a pump beam. The polarization state of the pump beam is set at a diagonal using the half-wave plate (HWP) and a quarter-wave plate (QWP). A PPKTP crystal with a poling period of $\Lambda=3.425~\mu m$ is quasi-phase-matched for the noncollinear type - 0 SPDC process.  

As given in the schematic diagram of the experimental set-up in Fig.\ref{QST}, the diagonally polarized pump beam enters the polarization Sagnac interferometer, consisting of a dual-polarizing beamsplitter (DPBS), a dual-wavelength half-waveplate (DHWP), and a pair of dual-band mirrors. The type-0 PPKTP crystal is kept inside the temperature oven set to $26.5^\circ C$ and placed at the balanced position of the interferometer. The pump beam is focused onto the centre of the PPKTP crystal using a plano-convex lens with $40~cm$ focal length. The DPBS splits the diagonally polarized pump beam into $|H_p\rangle$ and $|V_p\rangle$ components, which travel in the counter-clockwise (CCW) and clockwise (CW) directions, respectively. The horizontally polarized pump $|H_p\rangle$ undergoes a unitary evolution in a DHWP set to $45^\circ$ and becomes vertically polarized $|V_p\rangle$. This $|V_p\rangle$ interacts with the PPKTP to produce a signal-idler pair with a central wavelength of $\lambda_s=\lambda_i=810~nm$ traveling in the CCW direction. The joint polarization state of the signal-idler pair is described by $|V_s\rangle \otimes |V_i\rangle$. Similarly, the $|V_p\rangle$ in the CW direction interacts with a PPKTP crystal to produce a signal-idler pair with a polarisation state $|V_s\rangle \otimes |V_i\rangle$. The joint quantum state is transformed to $|H_s\rangle \otimes |H_i\rangle$ after the DHWP. The signal-idler pair $|H_s\rangle \otimes |H_i\rangle$ traveling in the CW direction and the $|V_s\rangle \otimes |V_i\rangle$ traveling in the CCW direction are combined at the DPBS and exit through the same output port. Ideally, the output two-photon polarization state is given by

\begin{equation}
|\phi^+\rangle = \frac{1}{\sqrt{2}} (|H_s\rangle \otimes |H_i\rangle + |V_s\rangle \otimes |V_i\rangle)
\end{equation}

The Bell state $|\phi^-\rangle = \frac{1}{\sqrt{2}} (|H_s\rangle \otimes |H_i\rangle - |V_s\rangle \otimes |V_i\rangle)$ generated by translating the PPKTP crystal from the balanced position of the interferometer by $122~\mu m$, following the procedure described by Thik et al.,\cite{Thik_2026}. The remaining two Bell states $|\psi^+\rangle = \frac{1}{\sqrt{2}} (|H_s\rangle \otimes |V_i\rangle + |V_s\rangle \otimes |H_i\rangle)$ and $|\psi^-\rangle=\frac{1}{\sqrt{2}} (|H_s\rangle \otimes |V_i\rangle - |V_s\rangle \otimes |H_i\rangle)$ using a HWP kept at an angle $45^\circ$ in the idler arm. Therefore, following the procedure given in Thik et al.~\cite{Thik_2026}, all four Bell states are generated with a single Sagnac interferometer. The quantum states of these four Bell states are measured using a pair of polarizers and the QWP combination in the signal and idler arms. The signal and idler photons are collected using collimators and coupled into single-mode fibres (SMFs), which are connected to single-photon avalanche diodes (SPADs) for photon detection. The SPAD outputs are connected with a coincidence counter unit. The density matrices of these four Bell states are measured using the standard QST and the tQST processes. The comparison results are given in the results section. 

\begin{figure}[htbp]
    \centering
    \begin{subfigure}{0.48\linewidth}
        \centering
        \includegraphics[width=\linewidth]{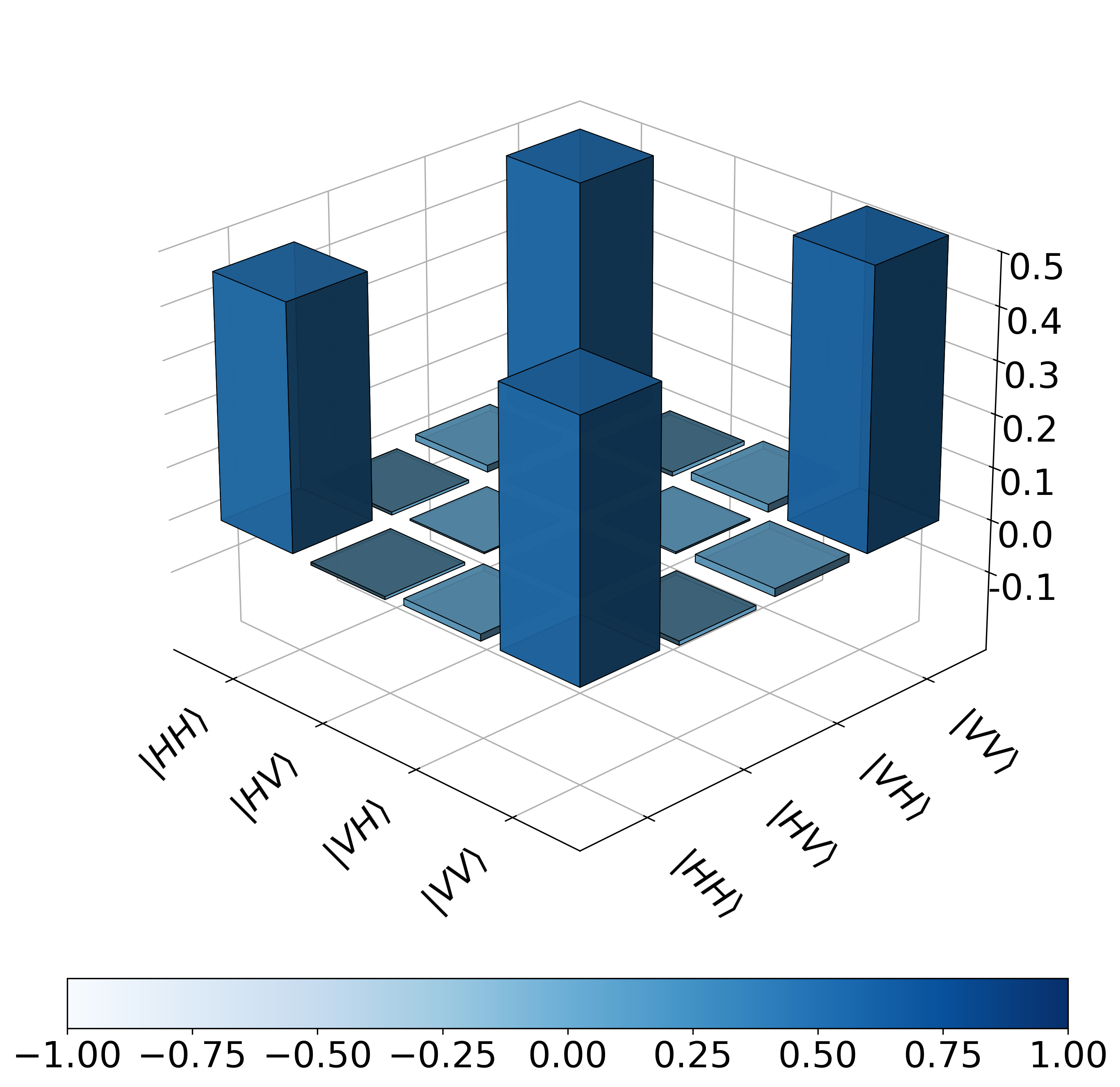}
        \caption{Real part of $\hat{\rho}_{sQST}$}
        \label{fig:real_qst}
    \end{subfigure}
    \hfill
    \begin{subfigure}{0.48\linewidth}
        \centering
        \includegraphics[width=\linewidth]{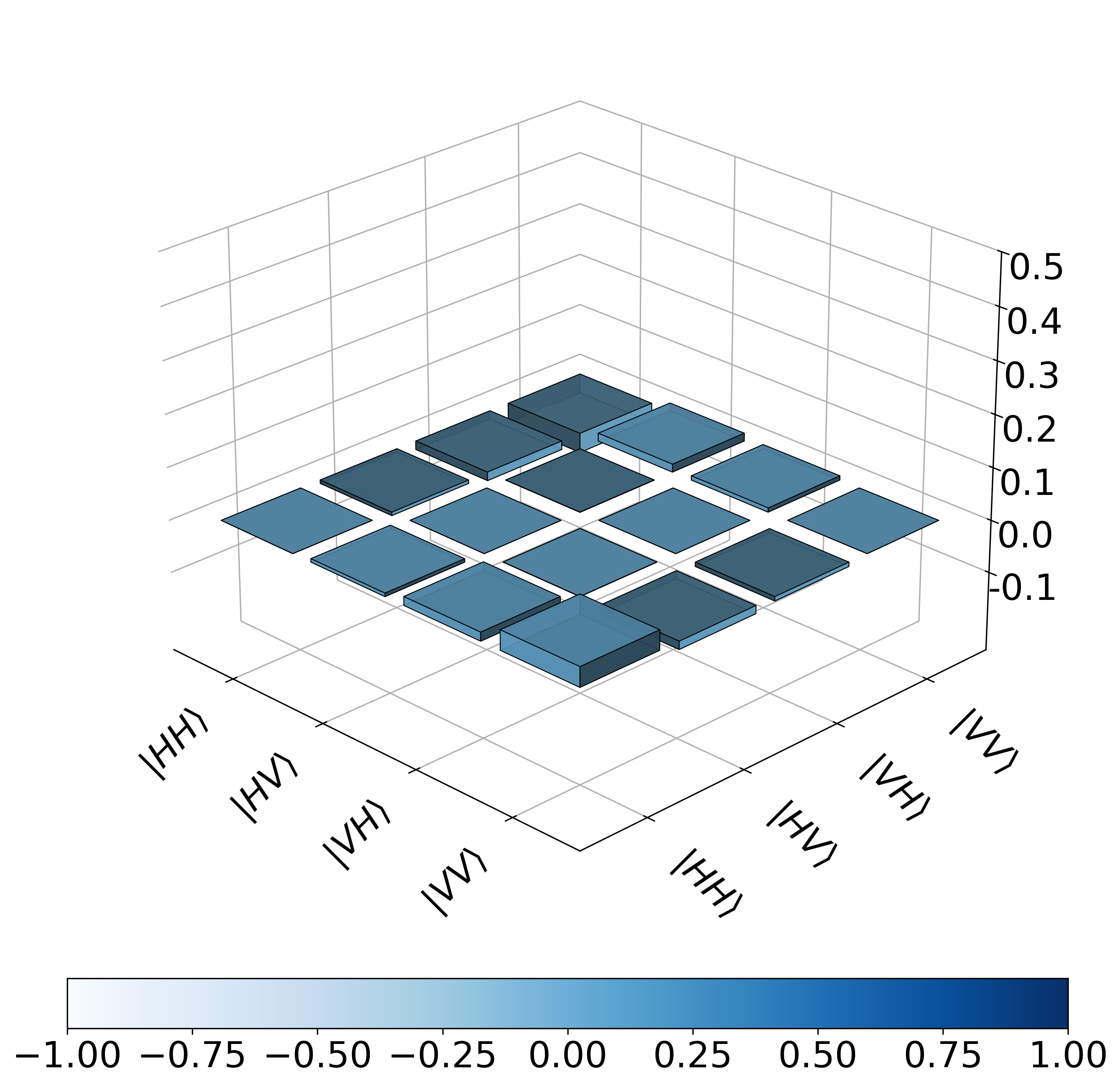}
        \caption{Imaginary part of $\hat{\rho}_{sQST}$}
        \label{fig:imag_qst}
    \end{subfigure}

    \vspace{1em}

    \begin{subfigure}{0.48\linewidth}
        \centering
        \includegraphics[width=\linewidth]{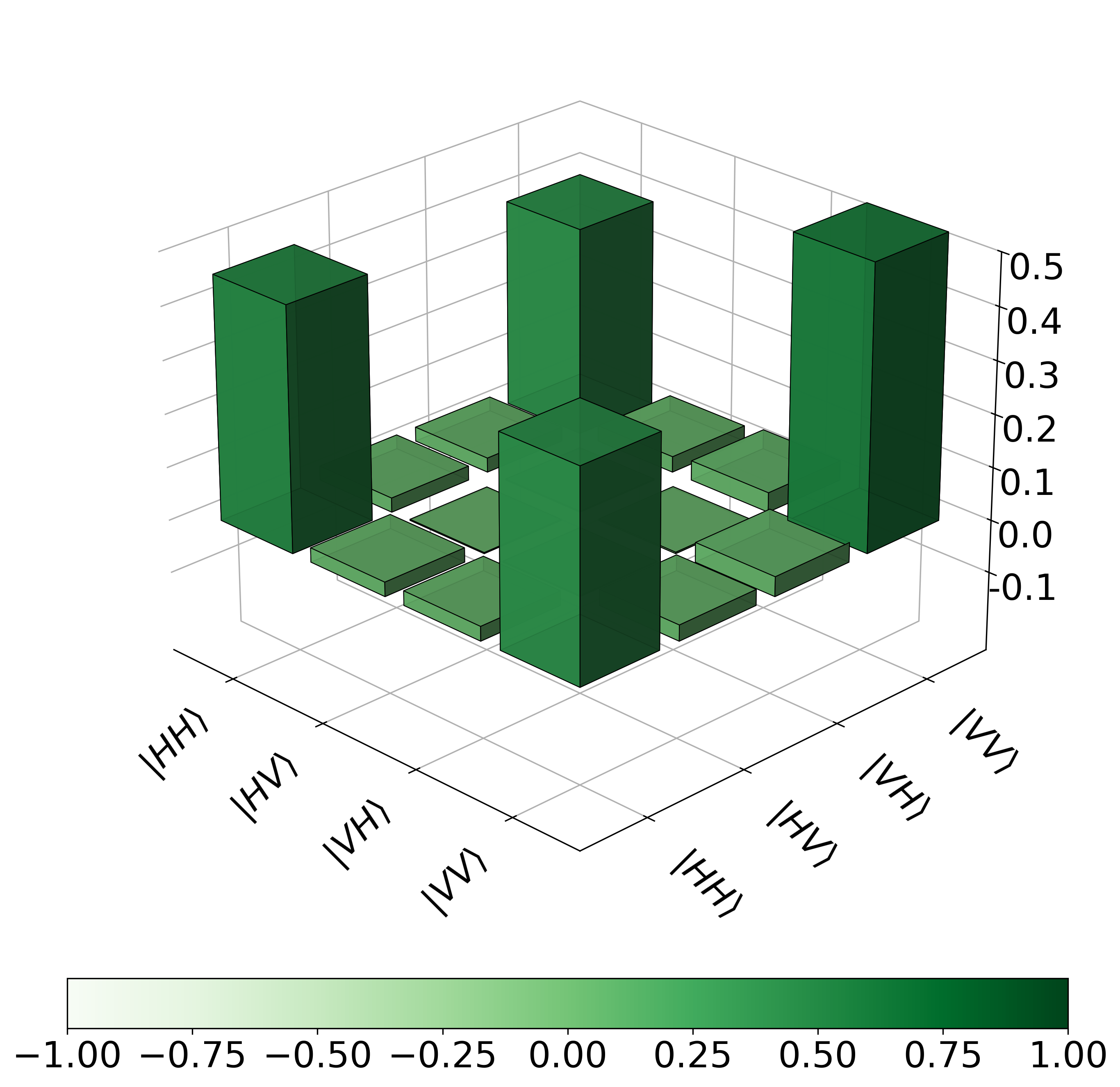}
        \caption{Real part of $\hat{\rho}_{tQST}$}
        \label{fig:real_tqst}
    \end{subfigure}
    \hfill
    \begin{subfigure}{0.48\linewidth}
        \centering
        \includegraphics[width=\linewidth]{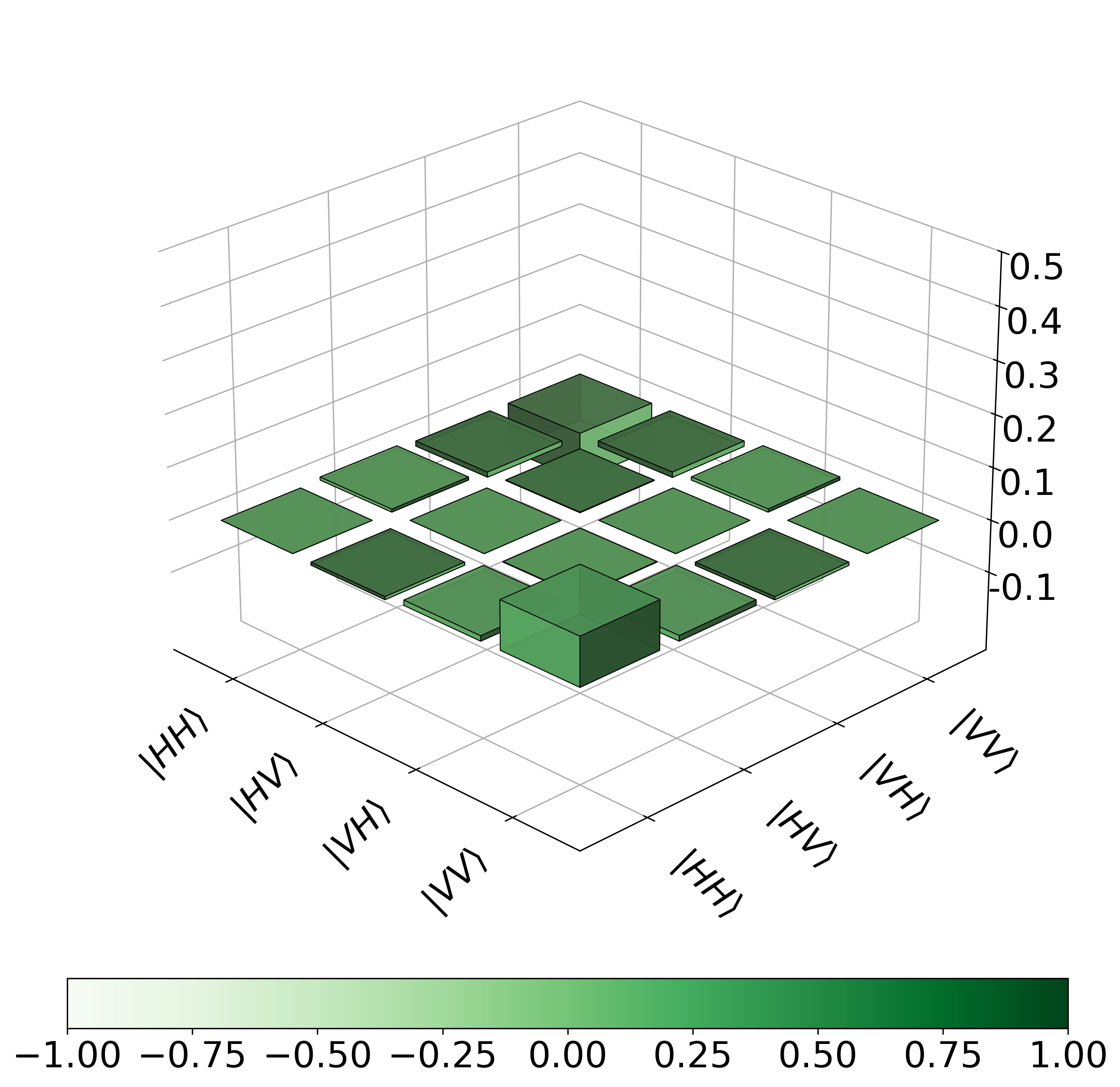}
        \caption{Imaginary part of $\hat{\rho}_{tQST}$}
        \label{fig:imag_tqst}
    \end{subfigure}

    \caption{\justifying Reconstructed density matrix of the $|\Phi^+\rangle$ state: real and imaginary parts from sQST (blue) and tQST (green).}
    \label{fig:rho_all}
\end{figure}

\begin{figure}[htbp]
    \centering
    \includegraphics[
    width=1\linewidth,
    height=6.5cm
]{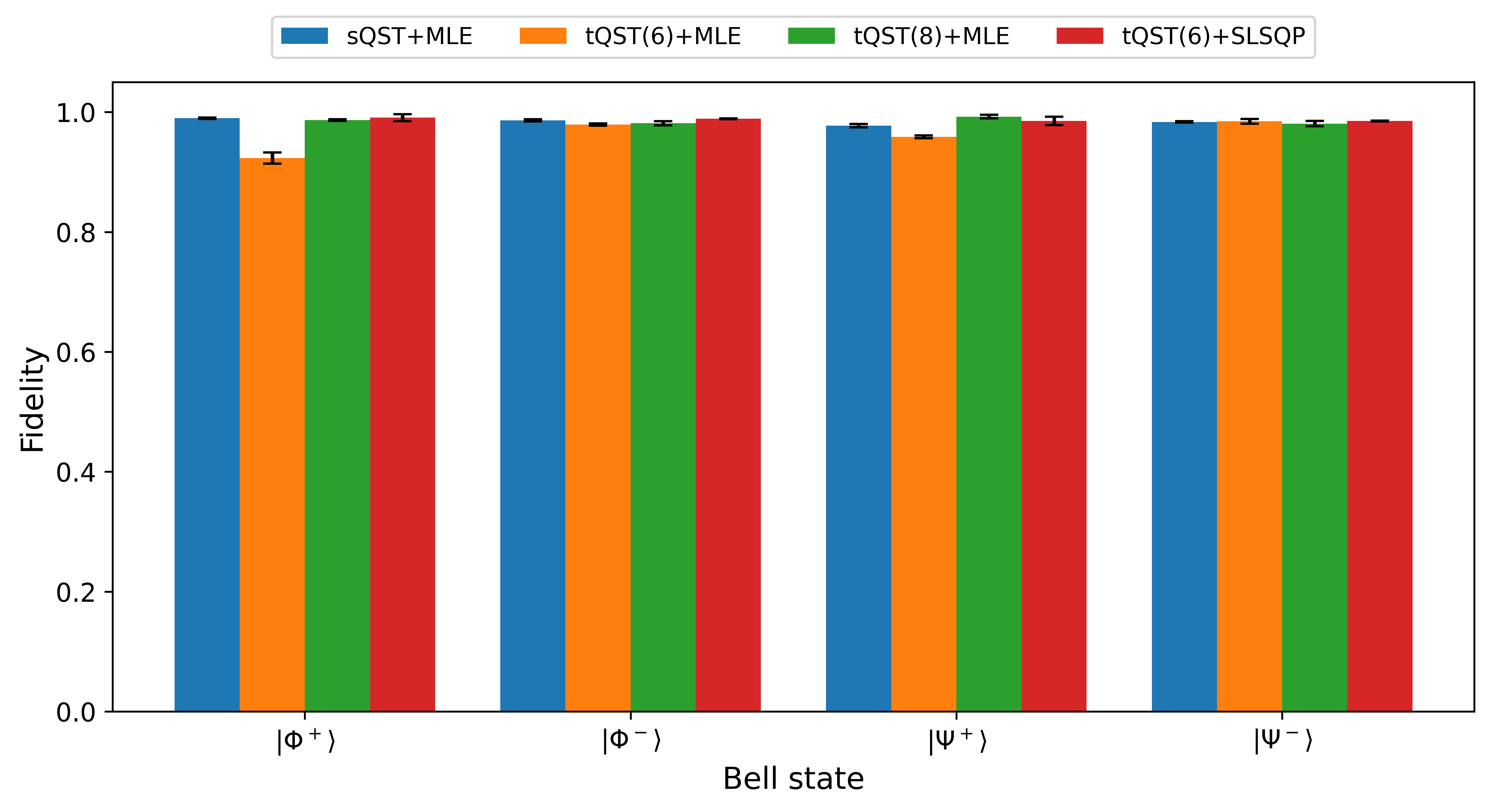}

\caption{\justifying
Comparison of the fidelity of all four
Bell states reconstructed using standard quantum state tomography with maximum-likelihood estimation (sQST + MLE, blue),
threshold quantum state tomography with six projective measurements and MLE (tQST (6) + MLE, orange), threshold quantum state tomography with eight projective measurements and MLE (tQST (8) + MLE, green), and threshold quantum state tomography with six projective
measurements and sequential least-squares programming (tQST (6) + SLSQP, red).
The bars represent the mean values, while the error bars indicate the corresponding standard deviations obtained from repeated measurements.}

    \label{fig:Fidelity_comparison}
\end{figure}

\begin{figure}[htbp]
    \centering
    \includegraphics[
    width=1\linewidth,
    height=6.5cm
]{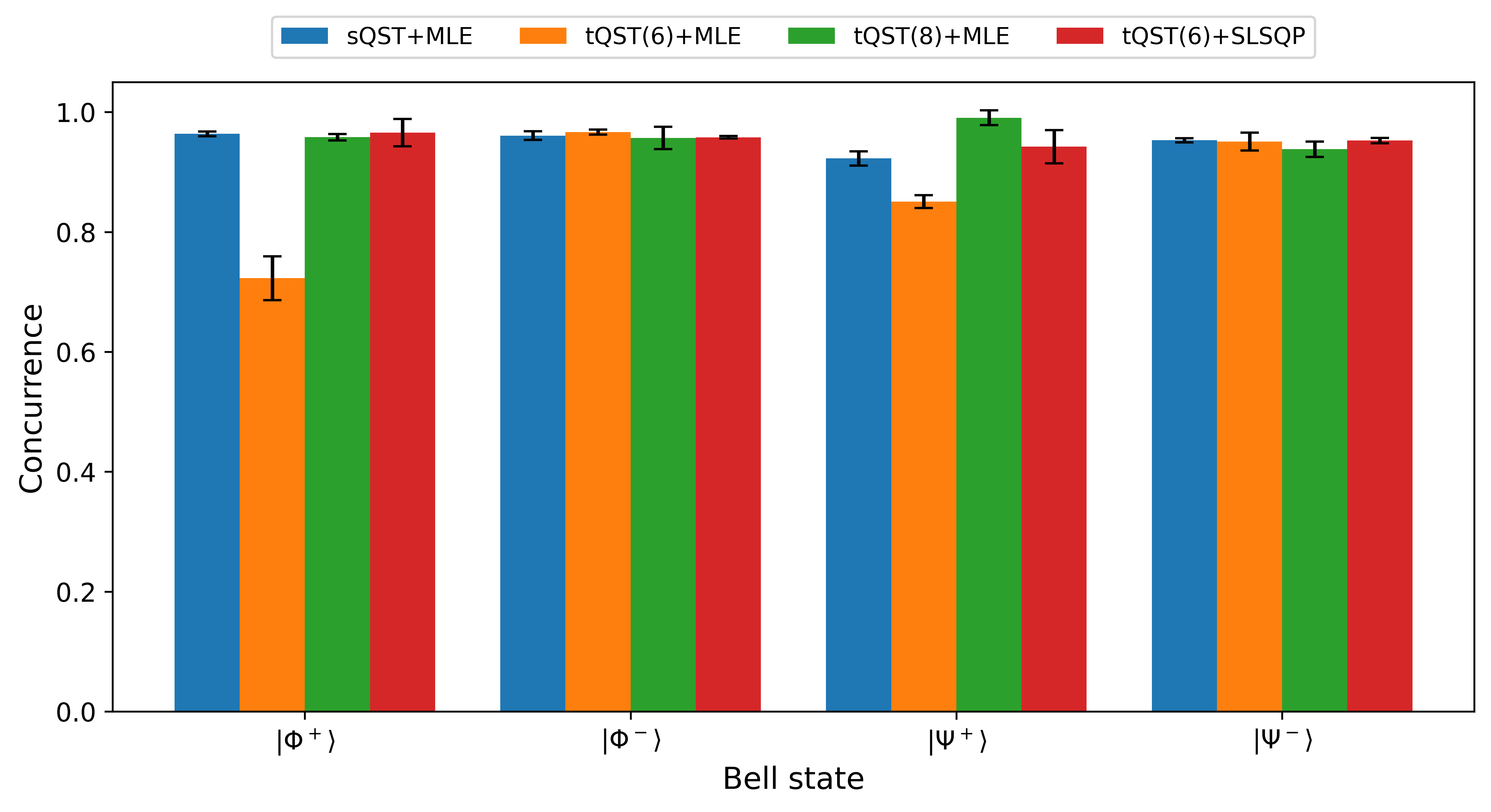}

\caption{\justifying
Comparison of the concurrence of all four
Bell states reconstructed using standard quantum state tomography
with maximum-likelihood estimation (sQST + MLE, blue),
threshold quantum state tomography with six projective
measurements and MLE (tQST (6) + MLE, orange),
threshold quantum state tomography with eight projective
measurements and MLE (tQST (8) + MLE, green),
and threshold quantum state tomography with six projective
measurements and sequential least-squares programming
(tQST (6) + SLSQP, red).
The bars represent the mean values, while the error bars
indicate the corresponding standard deviations obtained
from repeated measurements.}
\label{fig:Concurrence_comparison}
\end{figure}

\begin{figure}[htbp]
    \centering
    \includegraphics[
    width=1\linewidth,
    height=6.5cm
]{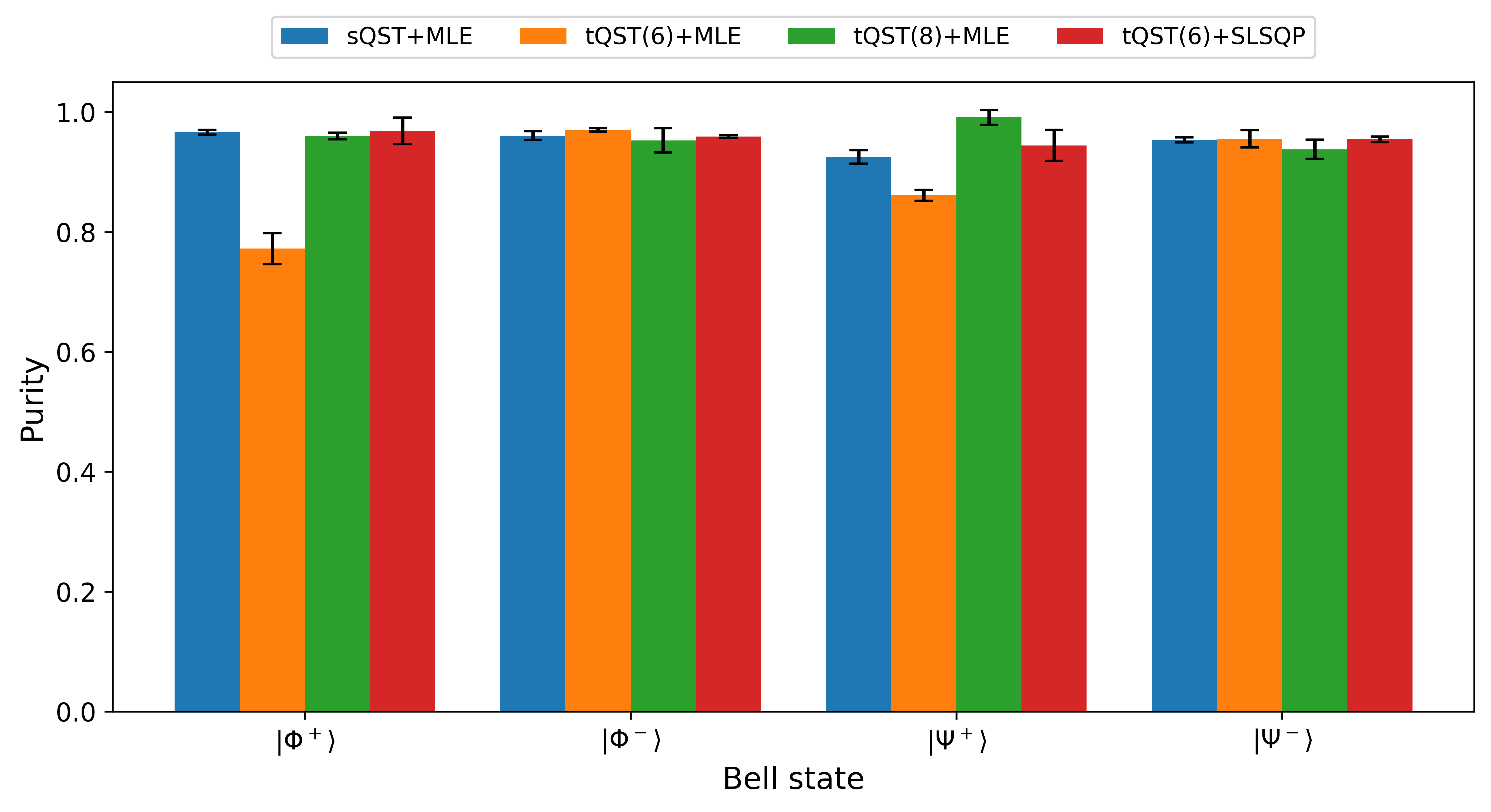}

\caption{\justifying
Comparison of the purity of all four
Bell states reconstructed using standard quantum state tomography
with maximum-likelihood estimation (sQST + MLE, blue),
threshold quantum state tomography with six projective
measurements and MLE (tQST (6) + MLE, orange),
threshold quantum state tomography with eight projective
measurements and MLE (tQST (8) + MLE, green),
and threshold quantum state tomography with six projective
measurements and sequential least-squares programming
(tQST (6) + SLSQP, red).
The bars represent the mean values, while the error bars
indicate the corresponding standard deviations obtained
from repeated measurements.}
\label{fig:Purity_comparison}
\end{figure}

\section{Results}
In this section, we present our experimental results for the threshold quantum state tomography (tQST) and compare it with the standard QST technique. We compare the characteristics of the two-qubit polarization-entangled states, such as fidelity, purity and concurrence, evaluated using the standard QST and the tQST process.

Quantum state reconstruction using standard QST requires sixteen projective measurements~\cite{Altepeter2004}. These projectors can be realized using a combination of a QWP and a polarizer in the signal and idler arms given in Fig.~\ref{QST}.

 As a starting point, the diagonal entries of the density operators are measured by projective measurements represented by $\{\hat{P}_{HH},\hat{P}_{HV},\hat{P}_{VH},\hat{P}_{VV}\}$, where the projector $\hat{P}_{ij}=|ij\rangle \langle ij |$ and $i, ~j \in \{H, V\}$. These projective measurements yield the probabilities associated with the standard basis states and they are related to the diagonal entries of the density matrix as, $\{\rho_{11}=p_{HH}, \rho_{22}=p_{HV}, \rho_{33}=p_{VH}, \rho_{44}=p_{VV}\}$. These probabilities $p_{ij}$ are associated with the projection operators $\hat{P}_{ij}$ and the density operators $\hat{\rho}$ as,
\begin{equation}
    p_{ij} = Tr(\hat{P}_{ij} \hat{\rho})
\end{equation}

The half-wave plate (HWP) in the pump beam path is
oriented at $22.5^\circ$ with respect to the horizontal
polarization axis to prepare the diagonally polarized
pump beam. The PPKTP crystal is positioned at the
balanced position of the Sagnac interferometer to
generate the $|\phi^+\rangle$ Bell state. The pump
power is maintained at $100~\mu\mathrm{W}$.
Bandpass filters centered at $810~\mathrm{nm}$,
each with a bandwidth of $10~\mathrm{nm}$, are
employed in both the signal and idler arms to
suppress unwanted spectral components.

The polarization states of the signal and idler
photons are analyzed using a combination of a
quarter-wave plate (QWP) and a polarizer in
each arm. For each projective measurement,
coincidence counts are recorded with an
integration time of $1~\mathrm{s}$, and the
measurement is repeated 20 times to obtain
20 experimental datasets. The experimentally
measured projection probabilities are reported
as the mean values obtained from these
20 acquisitions.

Using the projective measurements associated
with the computational basis, the mean diagonal
populations of the density matrix are obtained as

\begin{align}
    \rho_{11} & = 0.46143 \\
    \rho_{22} & = 0.00309 \\ 
    \rho_{33} & = 0.00323 \\
    \rho_{44} & = 0.53225
\end{align}

From Eqn (~\ref{eq:Gini1}), the calculated value of the Gini index is 
\begin{equation}
    GI(\hat{\rho}) = 0.5116
\end{equation}
From the Gini index $GI(\hat{\rho})$, the calculated value of the threshold value is
\begin{equation}
    t = 0.1705
    \label{thre}
\end{equation}

Now, the upper bounds on the off-diagonal coherence magnitudes are calculated from the diagonal as $B_{ij}=\sqrt{\rho_{ii}\rho_{jj}}$. The calculated values of the upper limits are as follows,

\[
\begin{aligned}
|B_{12}| = |B_{21}| &= 0.0373\\
|B_{13}| = |B_{31}| &= 0.0383 \\
|B_{14}| = |B_{41}| & = \textbf{0.4955} \\
|B_{23}| = |B_{32}| & = 0.0031 \\
|B_{24}| = |B_{42}| & = 0.0401 \\
|B_{34}| = |B_{43}| & = 0.0412
\end{aligned}
\]
Of these twelve off-diagonal entries, only two entries are above the threshold value given in Eqn (~\ref{thre}),
\begin{equation}
    |B_{14}| = |B_{41}| = 0.4955 \geq 0.1705
\end{equation}
The remaining ten off-diagonal entries are less than the threshold value, $t = 0.1705$.

Now, the projection operators associated with the off-diagonal entries $\rho_{14}$ and $\rho_{41}$ are identified using Eqn (\ref{eq:proj_selection}) as given in the table~\ref{tab:Projectors_matrix}. As these two matrix elements are complex conjugates of each other, finding one element gives information about the other, so we need to find the projectors for only one element, $\rho_{14}$. Using table~\ref{tab:Projectors_matrix}, the projector for the real part of $\rho_{14}$ is $\hat{P}_{DD} = |DD\rangle\langle DD|$, and the projector for the imaginary part of $\rho_{14}$ is $\hat{P}_{DR}=|DR\rangle\langle DR|$. Thus, the projective measurements onto the
two-qubit states $|DD\rangle$ and $|DR\rangle$
provide information about the real and imaginary
components of the selected off-diagonal
element $\rho_{14}$, respectively.
The corresponding measurement probabilities,
together with the experimentally determined
diagonal populations, are used to constrain
the density-matrix reconstruction. Hence, only six projective measurements are necessary for tQST, that is, four projective measurements for diagonal measurements and two for off-diagonal measurements. This is much less than the 16 measurements required for standard QST.

\begin{figure}[htbp]
    \centering
    \includegraphics[
    width=1\linewidth,
    height=6.5cm
]{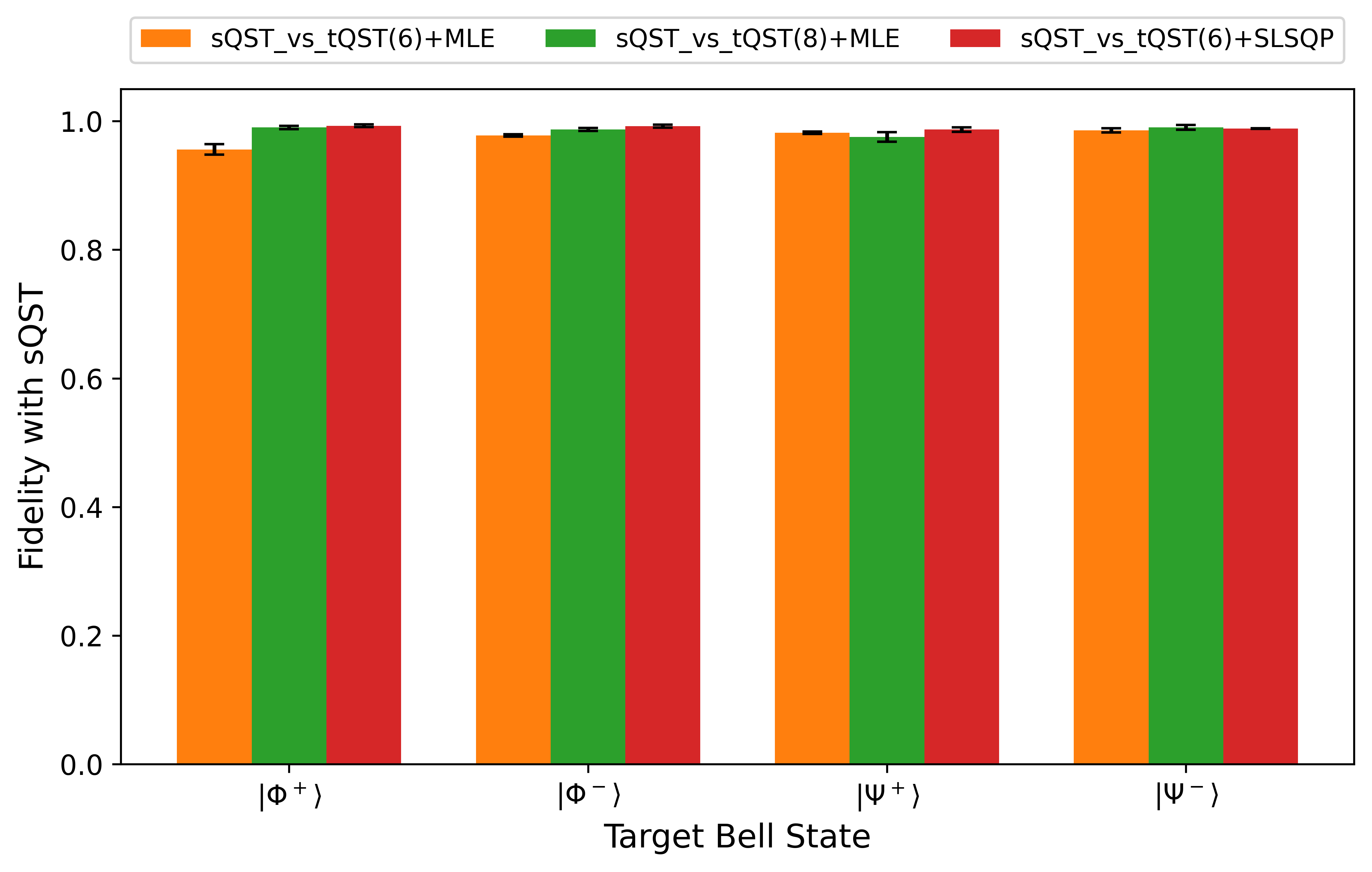}

\caption{\justifying Comparison of the fidelity of all four
Bell states reconstructed through
threshold quantum state tomography with six projective
measurements and MLE (tQST (6) + MLE, orange),
threshold quantum state tomography with eight projective
measurements and MLE (tQST (8) + MLE, Green),
and threshold quantum state tomography with six projective
measurements and sequential least-squares programming
(tQST (6) + SLSQP, Red)  with the density matrix reconstructed through the standard quantum state tomography
with maximum-likelihood estimation.
The bars represent the mean values, while the error bars
indicate the corresponding standard deviations obtained
from repeated measurements.}

    \label{fig:Fid_comparison-sQST}
\end{figure}

\begin{table*}[t]
    \centering
   \caption{\justifying Experimentally measured probabilities $p_{HH}$, $p_{HV}$, $p_{VH}$, and $p_{VV}$ corresponding to the diagonal elements of the density matrices of all four Bell states. The Gini index $GI(c)$ and the corresponding threshold $t$ are calculated from the measured diagonal populations using Eqs.~(\ref{eq:Gini1}) and (\ref{Eq:thresholdE2}), respectively.}
    \label{tab:threshold}
    \begin{ruledtabular}
    \begin{tabular}{|ccccccc}
        Quantum state
        & \multicolumn{4}{c}{Probabilities}
        & Gini index $GI(\rho)$
        & Threshold $t$ \\
        \cline{2-5}
        & $p_{HH}$ & $p_{HV}$ & $p_{VH}$ & $p_{VV}$
        & & \\
        \hline
        $|\phi^+\rangle$ & 0.46143 & 0.00309 & 0.00323 & 0.53225 &  0.5116 & 0.1705 \\
        \hline
        $|\phi^-\rangle$ & 0.46958 & 0.01038 & 0.01059 & 0.50945 & 0.4892 & 0.1631 \\
        \hline
        $|\psi^+\rangle$ & 0.00551 & 0.51168 & 0.47746 & 0.00535 & 0.4979 & 0.1659 \\
        \hline
        $|\psi^-\rangle$ & 0.00683 & 0.51825 & 0.46875 & 0.00617 & 0.4996 & 0.1665 \\
    \end{tabular}
    \end{ruledtabular}
    \label{tab:Allfour_threshold}
\end{table*}

\begin{table*}[t]
    \centering
    \caption{\justifying Selection of the projective measurements required to reconstruct the threshold-selected off-diagonal density-matrix elements for each Bell state. The upper bound on the magnitude of each off-diagonal element is given by $B_{ij}=\sqrt{\rho_{ii}\rho_{jj}}$, where $\rho_{ii}$ and $\rho_{jj}$ are determined from the experimentally measured diagonal populations. Values of $B_{ij}$ satisfying the threshold criterion $B_{ij}\geq t$ are highlighted in bold. The projectors associated with the selected off-diagonal elements are identified using Table~\ref{tab:Projectors_matrix} and listed in the final column.}
    \label{tab:threshold2}
    \begin{ruledtabular}
    \begin{tabular}{ccccccccc}
        Quantum state
        & Threshold $t$
        & $|B_{12}|$ 
        & $|B_{13}|$ 
        & $|B_{14}|$ 
        & $|B_{23}|$
        & $|B_{24}|$
        & $|B_{34}|$
        & Required Projectors \\ [3pt]
        \hline
        $|\phi^+\rangle$ & 0.1705 & $0.0373$ & $0.0383$ & $\textbf{0.4955}$ & $0.0031$ & 0.0401 & 0.0412 & $\hat{P}_{DD}, \hat{P}_{DR}$ \\ [3pt]
        \hline 
        $|\phi^-\rangle$ & 0.1631 & $0.0697$ & $0.0705$ & $\textbf{0.4891}$ & $0.0105$ & 0.0726 & 0.0734 & $\hat{P}_{DD}, \hat{P}_{DR}$ \\ [3pt]
        \hline
        $|\psi^+\rangle$ & 0.1659 & $0.0530$ & $0.0512$ & $0.0054$ & $\textbf{0.4942}$ & 0.0522 & 0.0504 & $\hat{P}_{RR}, \hat{P}_{RD}$ \\ [3pt]
        \hline 
        $|\psi^-\rangle$ & 0.1665 & $0.0594$ & $0.0565$ & $0.0065$ & $\textbf{0.4928}$ & 0.0564 & 0.0537 & $\hat{P}_{RR}, \hat{P}_{RD}$ \\ [3pt]
    \end{tabular}
    \end{ruledtabular}
    \label{tab:Allfour_projectors}
\end{table*}

\begin{table}[htbp]
\centering
\caption{\justifying Comparison of fidelities between standard QST and tQST for the four Bell states.}
\resizebox{\columnwidth}{!}{%
\begin{tabular}{|c|c|c|c|}
\hline
\textbf{Sr. No} & \textbf{Bell State} & \textbf{Fidelity(\(\rho^{\text{sQST}}\))} & \textbf{Fidelity(\(\rho^{\text{tQST}}\))} \\
\hline
1. & \(|\phi\rangle^+ = \frac{1}{\sqrt{2}} (|HH\rangle + |VV\rangle)\) & 0.9897 & 0.9235 \\
\hline
2. & \(|\phi\rangle^- = \frac{1}{\sqrt{2}} (|HH\rangle - |VV\rangle)\) & 0.9862 & 0.9792 \\
\hline
3. & \(|\psi\rangle^+ = \frac{1}{\sqrt{2}} (|HV\rangle + |VH\rangle)\) & 0.9774 & 0.9586 \\
\hline
4. & \(|\psi\rangle^- = \frac{1}{\sqrt{2}} (|HV\rangle - |VH\rangle)\) & 0.9836 & 0.9847 \\
\hline
\end{tabular}%
}
\label{tab:bell_fidelities}
\end{table}

\begin{table}[htbp]
\centering
\caption{\justifying Concurrence and purity of all four Bell states calculated from the density matrices reconstructed using threshold quantum state tomography (tQST).}
\resizebox{\columnwidth}{!}{%
\begin{tabular}{|c|c|c|c|}
\hline
\textbf{Sr. No} & \textbf{Bell State} & \textbf{Concurrence} & \textbf{Purity} \\
\hline
1. & \(|\phi\rangle^+ = \frac{1}{\sqrt{2}} (|HH\rangle + |VV\rangle)\) & 0.7228 & 0.7722 \\
\hline
2. & \(|\phi\rangle^- = \frac{1}{\sqrt{2}} (|HH\rangle - |VV\rangle)\) & 0.9667  & 0.9704 \\
\hline
3. & \(|\psi\rangle^+ = \frac{1}{\sqrt{2}} (|HV\rangle + |VH\rangle)\) & 0.8505 & 0.8613 \\
\hline
4. & \(|\psi\rangle^- = \frac{1}{\sqrt{2}} (|HV\rangle - |VH\rangle)\) & 0.9509 & 0.9553 \\
\hline
\end{tabular}%
}
\label{tab:stateproperties}
\end{table}

As given in table \ref{tab:Projectors_matrix}, the projective measurements associated with the
real and imaginary parts of the off-diagonal
element $\rho_{14}$ are given by
\begin{equation}
\hat{P}_{14}^{\mathrm{re}} =
\hat{P}_{DD} = |DD\rangle\langle DD|,
\end{equation}
and
\begin{equation}
\hat{P}_{14}^{\mathrm{im}} =
\hat{P}_{DR} = |DR\rangle\langle DR|.
\end{equation}

The corresponding projection probabilities are expressed as,
\begin{align}
    p_{DD} &= Tr(\hat{\rho}\hat{P}_{DD}) \\
    p_{DR} &= Tr(\hat{\rho}\hat{P}_{DR})
\end{align}

Using the outcomes of these six projective measurements,
the density matrix of the experimentally generated
Bell state $|\phi^+\rangle$ is estimated following
the tQST procedure introduced by Binosi
\textit{et al.}~\cite{10.1063/5.0219143}.
Statistical fluctuations and experimental imperfections
can lead to an unphysical density matrix in direct
reconstruction procedures. Therefore, we employ
maximum-likelihood estimation (MLE)
~\cite{Altepeter2004} to obtain a physically valid
density matrix consistent with the experimental
measurement data. The MLE procedure ensures that the reconstructed density matrix satisfies the conditions of Hermiticity, unit trace, and positive semidefiniteness. Thus, the density matrix of the
Bell state $|\phi^+\rangle$ is reconstructed using
only six projective measurements, compared with
the sixteen measurements employed in conventional
QST. For each projective measurement, 20 independent
experimental single and concidence counts acquired, and
the average values of the corresponding projection probabilities calculated. The density matrix is then
reconstructed using these mean probabilities.
The fidelity, concurrence, and purity are
subsequently calculated from the reconstructed
density matrix. The real and imaginary parts of the density matrix reconstructed for the Bell state
$|\phi^+\rangle$ are presented in Fig~\ref{fig:rho_all}.

For the same Bell state, the density operator is reconstructed using the standard QST~\cite{Altepeter2004}, which requires the sixteen projective measurements. The real and imaginary parts of the density matrix reconstructed through the standard QST process are given in Fig.~\ref{fig:real_qst} and ~\ref{fig:imag_qst}.

We apply the tQST protocol to all four Bell states
generated using the experimental setup described in Section III. The four Bell states are prepared by
controlling the relative phase between the two
counterpropagating SPDC processes through translation
of the PPKTP crystal within the Sagnac interferometer
and by introducing a HWP oriented
at $45^\circ$ with respect to the horizontal
polarization axis in the idler arm. The measured values of the diagonal entries of four Bell states are given in the table ~\ref{tab:Allfour_threshold}. The corresponding Gini index $GI(\hat{\rho})$ and the threshold value $t$ are calculated using Eqn (\ref{eq:Gini1}) and Eqn (\ref{Eq:thresholdE2}). Based on these threshold values, the projective measurements selected for the reconstruction of each Bell state are listed in Table~\ref{tab:Allfour_projectors}.

The fidelity between the experimentally reconstructed
quantum state $\hat{\rho}_{\mathrm{exp}}$ and the
ideal target Bell state with density matrix
$\hat{\rho}_{\mathrm{ideal}}$ is defined as
\begin{equation}
F(\hat{\rho}_{\mathrm{ideal}},
\hat{\rho}_{\mathrm{exp}})
=
\operatorname{Tr}
\left[
\sqrt{
\sqrt{\hat{\rho}_{\mathrm{ideal}}}
\hat{\rho}_{\mathrm{exp}}
\sqrt{\hat{\rho}_{\mathrm{ideal}}}
}
\right].
\label{eq:fidelity}
\end{equation}

To quantify the degree of two-qubit entanglement, we also evaluate the concurrence~\cite{Wootters1998}, defined as
\begin{equation}
C(\hat{\rho}) = \max\left\{0, \lambda_1 - \lambda_2 - \lambda_3 - \lambda_4\right\},
\end{equation}
where $\lambda_1 \ge \lambda_2 \ge \lambda_3 \ge \lambda_4$ are the square roots of the eigenvalues, in decreasing order, of the non-Hermitian matrix $\hat{\rho}(\hat{\sigma}_y \otimes \hat{\sigma}_y)\hat{\rho}^*(\hat{\sigma}_y \otimes \hat{\sigma}_y)$, with $\hat{\sigma}_y$ denoting the Pauli-$Y$ operator and $\hat{\rho}^*$ representing the complex conjugate of $\hat{\rho}$ in the computational basis.

Table~\ref{tab:bell_fidelities} compares the fidelities of the quantum states reconstructed using standard quantum state tomography (sQST) and threshold quantum state tomography (tQST). The density matrices reconstructed using sQST exhibit fidelities exceeding $0.9774$ with respect to the corresponding ideal Bell states. In comparison, the six-projector tQST reconstruction yields fidelities of $0.9235$ and $0.9586$ for $|\Phi^+\rangle$ and $|\Psi^+\rangle$, respectively, while higher fidelities of $0.9792$ and $0.9847$ are obtained for $|\Phi^-\rangle$ and $|\Psi^-\rangle$, respectively. Although all four Bell states exhibit fidelities above $0.9235$, the relatively lower values obtained for $|\Phi^+\rangle$ and $|\Psi^+\rangle$ indicate differences in the reconstruction performance. We further evaluated the concurrence and purity of the reconstructed states and observed comparatively lower values for $|\Phi^+\rangle$ than for the other Bell states as given in Table \ref{tab:stateproperties}.

The reduced reconstruction fidelity observed for certain Bell states may arise from two possible factors: (i) insufficient measurement information provided by the reduced set of projectors and (ii) ambiguities or limitations associated with the numerical reconstruction procedure. Since the six-projector measurement set is not informationally complete for an arbitrary two-qubit state, the reconstruction may depend on the constraints and parametrization employed during numerical estimation. These observations motivate a further investigation of the trade-off between the number of projective measurements and the accuracy of the reconstructed quantum states.

To investigate these discrepancies, we considered two complementary approaches. In the first approach, the number of projective measurements was increased from six to eight. In addition to the four projectors ${\hat{P}_{HH},\hat{P}_{HV},\hat{P}_{VH},\hat{P}_{VV}}$ used to determine the diagonal populations, we considered four additional projectors ${\hat{P}_{DD},\hat{P}_{DR},\hat{P}_{RR},\hat{P}_{RD}}$, instead of the two off-diagonal projectors employed in the original six-projector reconstruction. The selection of these measurements is motivated by the structure of the ideal Bell-state density matrices, in which the only nonzero independent off-diagonal coherence is $\rho_{14}$ for $|\Phi^\pm\rangle$ and $\rho_{23}$ for $|\Psi^\pm\rangle$. The two additional projective measurements provide further constraints on the reconstructed density matrix, thereby reducing the ambiguity associated with the limited measurement information. The eight-projector reconstruction yields improved fidelity, concurrence, and purity, particularly for the states that exhibited comparatively lower values with six measurements. 

In the second approach, we investigated a constrained reconstruction procedure based on sequential least-squares programming (SLSQP), while retaining the original six-projector measurement set. The numerical reconstruction provided in the original tQST implementation by Binosi et al.~\cite{10.1063/5.0219143} employs a general triangular density-matrix parametrization and minimizes a Gaussian-likelihood-based weighted-residual objective using the L-BFGS-B optimization algorithm~\cite{Byrd1995LBFGSB}. The triangular parametrization ensures Hermiticity and positive semidefiniteness through the construction of the density matrix as $\rho=T^\dagger T/\operatorname{Tr}(T^\dagger T)$, where $T$ is a complex triangular matrix.

\begin{table*}[t]
    \centering
    \caption{\justifying Comparison of the Fidelity, Purity and concurrence for all four Bell states reconstructed using standard quantum state tomography with maximum-likelihood estimation, threshold quantum state tomography with six projective measurements and MLE, threshold quantum state tomography with eight projective measurements and MLE, and threshold quantum state tomography with six projective measurements and sequential least-squares programming.}
    \label{tab:bell_state_metrics}
    \renewcommand{\arraystretch}{1.3}
    \setlength{\tabcolsep}{10pt}
    \begin{tabular}{@{} l l c c c c @{}}
        \toprule
        \textbf{State} & \textbf{Metric} &
        \textbf{sQST+MLE} &
        \textbf{tQST(6)+MLE} &
        \textbf{tQST(8)+MLE} &
        \textbf{tQST(6)+SLSQP} \\
        \midrule

        \multirow{3}{*}{$|\Phi^+\rangle$}
        & Fidelity
        & $0.9897 \pm 0.0010$
        & $0.9235 \pm 0.0093$
        & $0.9865 \pm 0.0015$
        & $0.9908 \pm 0.0059$ \\

        & Concurrence
        & $0.9637 \pm 0.0039$
        & $0.7228 \pm 0.0367$
        & $0.9580 \pm 0.0055$
        & $0.9657 \pm 0.0230$ \\

        & Purity
        & $0.9664 \pm 0.0038$
        & $0.7722 \pm 0.0259$
        & $0.9601 \pm 0.0056$
        & $0.9688 \pm 0.0221$ \\
        \midrule

        \multirow{3}{*}{$|\Phi^-\rangle$}
        & Fidelity
        & $0.9862 \pm 0.0020$
        & $0.9792 \pm 0.0019$
        & $0.9815 \pm 0.0035$
        & $0.9889 \pm 0.0007$ \\

        & Concurrence
        & $0.9608 \pm 0.0071$
        & $0.9667 \pm 0.0041$
        & $0.9569 \pm 0.0186$
        & $0.9580 \pm 0.0020$ \\

        & Purity
        & $0.9608 \pm 0.0074$
        & $0.9704 \pm 0.0027$
        & $0.9528 \pm 0.0204$
        & $0.9594 \pm 0.0019$ \\
        \midrule

        \multirow{3}{*}{$|\Psi^+\rangle$}
        & Fidelity
        & $0.9774 \pm 0.0030$
        & $0.9586 \pm 0.0024$
        & $0.9923 \pm 0.0030$
        & $0.9853 \pm 0.0070$ \\

        & Concurrence
        & $0.9227 \pm 0.0119$
        & $0.8505 \pm 0.0108$
        & $0.9906 \pm 0.0123$
        & $0.9422 \pm 0.0277$ \\

        & Purity
        & $0.9251 \pm 0.0110$
        & $0.8613 \pm 0.0091$
        & $0.9912 \pm 0.0124$
        & $0.9444 \pm 0.0259$ \\
        \midrule

        \multirow{3}{*}{$|\Psi^-\rangle$}
        & Fidelity
        & $0.9836 \pm 0.0012$
        & $0.9847 \pm 0.0040$
        & $0.9808 \pm 0.0046$
        & $0.9852 \pm 0.0008$ \\

        & Concurrence
        & $0.9529 \pm 0.0036$
        & $0.9509 \pm 0.0150$
        & $0.9381 \pm 0.0128$
        & $0.9524 \pm 0.0046$ \\

        & Purity
        & $0.9536 \pm 0.0042$
        & $0.9553 \pm 0.0144$
        & $0.9378 \pm 0.0161$
        & $0.9544 \pm 0.0046$ \\

        \bottomrule
    \end{tabular}
    \label{tab:comparison_table}
\end{table*}

In contrast, our SLSQP-based reconstruction employs a direct Hermitian parametrization, in which the diagonal populations and the real and imaginary components of the threshold-selected off-diagonal coherences are treated as optimization variables. The remaining off-diagonal coherences are explicitly set to zero. Hermiticity is ensured by the parametrization, while unit trace and positive semidefiniteness are imposed through equality and inequality constraints, respectively. A probability-weighted least-squares objective, closely related to that employed in the original tQST implementation, is minimized using the SLSQP algorithm.

An important distinction between the two procedures is the number of independent parameters considered during reconstruction. The general two-qubit density matrix contains $15$ independent real parameters, whereas the threshold-restricted Hermitian parametrization contains $3+2k$ independent real parameters after imposing the unit-trace condition, where $k$ denotes the number of retained independent complex off-diagonal coherence pairs. For a single retained coherence pair ($k=1$), the number of independent parameters is reduced from $15$ to $5$. 

Thus, although both reconstruction procedures employ the projective measurements selected through the tQST threshold criterion, they differ primarily in two aspects: (i) the parametrization and number of adjustable density-matrix elements and (ii) the numerical optimization algorithms used for reconstruction. Consequently, the improved reconstruction performance cannot be attributed solely to the SLSQP algorithm, but may also arise from the reduced number of independent parameters in the threshold-restricted density-matrix model.

To assess the performance of these approaches, we compared the fidelity, concurrence, and purity of all four Bell states reconstructed using four procedures: (i) conventional sQST with $16$ projective measurements and the MLE, (ii) tQST with six projective measurements and the MLE, (iii) tQST with eight projective measurements and the MLE, and (iv) tQST with six projective measurements using the constrained SLSQP-based reconstruction procedure. The fidelity, purity and concurrence values are given in Table~\ref{tab:comparison_table}. The histograms for the corresponding comparisons are presented in Figs.~\ref{fig:Fidelity_comparison}, \ref{fig:Concurrence_comparison}, and \ref{fig:Purity_comparison}, respectively.

We have also measured the fidelities of the density matrices of all four Bell states reconstructed through these three tQST procedures with the density matrices reconstructed using the sQST process. The comparison histogram for the fidelities of all four Bell states is given in Fig.~\ref{fig:Fid_comparison-sQST}.

The results indicate that incorporating additional projective measurements and employing the threshold-restricted constrained reconstruction procedure improves the agreement of the reconstructed states with the corresponding ideal Bell states, particularly for $|\Phi^+\rangle$ and $|\Psi^+\rangle$. The bars represent the mean values obtained from repeated measurements, while the error bars indicate the corresponding standard deviations. These observations demonstrate that, in addition to the number of projective measurements, the choice of density-matrix parametrization and numerical reconstruction procedure can influence the performance of measurement-reduced quantum state tomography.  A statistical comparison based on repeated measurements further shows that the standard deviations of the reconstructed fidelity, concurrence, and purity are not necessarily smaller for tQST than for standard QST, despite the reduced number of projective measurements.

\section{Conclusion}

In summary, we have experimentally investigated and compared threshold quantum state tomography (tQST) with conventional quantum state tomography (QST) for two-qubit polarization-entangled photons generated via the spontaneous parametric down-conversion (SPDC) process. We employed a type-0 phase-matched periodically poled potassium titanyl phosphate (PPKTP) crystal in a Sagnac interferometric configuration to generate polarization-entangled photon pairs. Using a single Sagnac-interferometer-based source, we generated all four Bell states. Switching between the different Bell states was achieved through controlled translation of the nonlinear crystal within the Sagnac interferometer and appropriate polarization manipulation using a HWP.

The four Bell states were characterized using
conventional quantum state tomography (sQST) with
16 projective measurements and threshold quantum
state tomography (tQST) with six projective
measurements. The density matrices reconstructed
using sQST were compared with those obtained
using three different tQST procedures:
(i) six projective measurements with
maximum-likelihood estimation (MLE),
(ii) eight projective measurements with MLE,
and (iii) six projective measurements with
a threshold-restricted reconstruction based
on sequential least-squares programming (SLSQP).

The reconstructed states exhibit good agreement
with the corresponding target Bell states, with
the reconstruction fidelity depending on the
number of projective measurements and the
state-estimation procedure. In particular,
the eight-projector MLE and six-projector
SLSQP-based reconstructions yield fidelities
exceeding 0.98 for the $|\psi^+\rangle$ and
$|\phi^+\rangle$ states.
The concurrence values calculated from the
reconstructed density matrices further confirm
the entanglement of the generated states.

Most importantly, tQST achieves a $62.5\%$
reduction in the number of projective
measurements compared with conventional QST
when six projectors are employed. These
results demonstrate the applicability of
tQST to polarization-entangled photon pairs
generated through SPDC and highlight the
trade-off between measurement efficiency
and reconstruction accuracy.

Two-qubit maximally entangled states are fundamental resources for a wide range of quantum-information applications, including quantum communication, quantum teleportation, superdense coding, quantum metrology and entanglement-based quantum key distribution~\cite{PhysRevLett.70.1895,PhysRevLett.69.2881,PhysRevLett.67.661,10.1116/5.0007577}. Accurate characterization of experimentally generated states is important for assessing their quality and validating their suitability for such applications. The reduced measurement requirement also provides a promising route toward more efficient and automated quantum-state characterization.

 In particular, multiqubit entangled states are important resources for quantum information protocols such as one-way quantum computing. The demonstrated reduction in measurement requirements for two-qubit polarization-entangled states, therefore, highlights the potential of tQST for resource-efficient characterization of larger photonic quantum systems.

\begin{acknowledgments}
The authors acknowledge the support from DIAT(DU) under the grant-in-aid program. The authors acknowledge useful discussions with Kanaka Raju Pandiri and Joseph Amalnathan (School of Quantum Technology, DIAT, Pune, India).
\end{acknowledgments}

\section*{Data Availability}

The experimental data supporting the findings of
this study, including the coincidence-count data
used for quantum-state reconstruction, are available
from the corresponding author upon reasonable request.
The MLE-based reconstruction was performed using the
original tQST implementation developed by Binosi
\textit{et al.}~\cite{10.1063/5.0219143}, which is
publicly available at
\url{https://github.com/liscidini-group/thresholdqst}.

\appendix

\section {The quantum state estimation using SLSQP and its comparison with the MLE process}

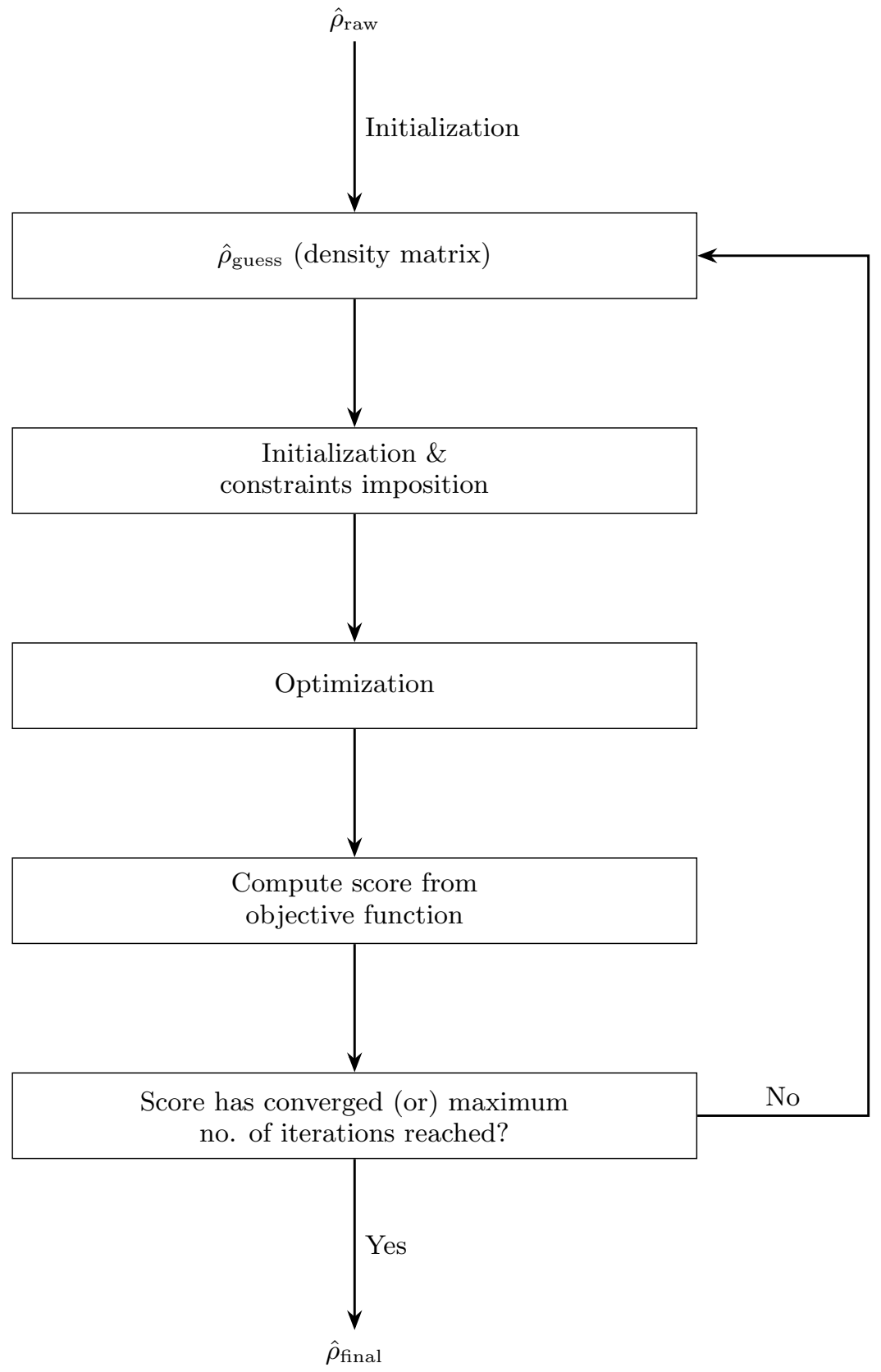
\begin{figure}[htbp]
\centering
\resizebox{0.8\linewidth}{!}{%
\begin{tikzpicture}[node distance=1.5cm, auto,
  box/.style={rectangle, draw, align=center, minimum width=8cm, minimum height=1cm},
  arrow/.style={-{Stealth}, thick}]

  \node (raw) {$\hat{\rho}_{\text{raw}}$};
  \node[box, below=2cm of raw] (guess) {$\hat{\rho}_{\text{guess}}$ (density matrix)};
  \node[box, below=of guess] (init) {Initialization \& \\ constraints imposition};
  \node[box, below=of init] (opt) {Optimization};
  \node[box, below=of opt] (score) {Compute score from \\ objective function};
  \node[box, below=of score] (decision) {Score has converged (or) maximum \\ no. of iterations reached?};
  \node[below=2cm of decision] (final) {$\hat{\rho}_{\text{final}}$};

  \draw[arrow] (raw) -- node[right] {Initialization} (guess);
  \draw[arrow] (guess) -- (init);
  \draw[arrow] (init) -- (opt);
  \draw[arrow] (opt) -- (score);
  \draw[arrow] (score) -- (decision);
  \draw[arrow] (decision) -- node[right] {Yes} (final);
  \draw[arrow] (decision.east) -- node[above] {No} ++(2,0) |- (guess.east);
\end{tikzpicture}%
}
\caption{\justifying General iterative workflow of a quantum state estimator, showing the progression from an unphysical raw matrix $\hat{\rho}_{\text{raw}}$ to a valid physical density matrix $\hat{\rho}_{\text{final}}$ through constraint imposition, optimization, and convergence checks.}
\label{fig:general_estimator}
\end{figure}

We have used two different quantum state estimation processes, namely the maximum likelihood estimator (MLE) and sequential least-squares programming (SLSQP). The MLE is widely used for QST in quantum information studies. In the present section, we discuss the quantum state estimation process using sequential least-squares programming (SLSQP). Let the initial matrix obtained from an $n$-qubit projection table, $\hat{\rho}_{\text{raw}}$, be unphysical. The estimator takes $\hat{\rho}_{\text{raw}}$ and iteratively maps it toward $\hat{\rho}_{\text{final}}$, which is a valid, physical density matrix that is as close to $\hat{\rho}_{\text{raw}}$ as possible. The intermediate density matrix evaluated during these iterative estimation steps is designated as $\hat{\rho}_{\text{guess}}$.

Fundamentally, the state estimator performs two primary functions, as illustrated in Fig.~\ref{fig:general_estimator}:
\begin{enumerate}
    \item \textbf{Initialization and Imposition of Constraints:} The estimator strictly enforces the required physical boundaries on $\hat{\rho}_{\text{guess}}$, forcing the Trace to equal 1 ($\text{Tr}(\hat{\rho}_{\text{guess}}) = 1$), ensuring the matrix remains Hermitian ($\hat{\rho}_{\text{guess}} = \hat{\rho}_{\text{guess}}^\dagger$), and imposing a positive semi-definite requirement where all eigenvalues are non-negative ($\lambda_i \ge 0$).
    \item \textbf{Parameter Tuning:} The estimator updates variables systematically, utilizing an objective function to compute a score and an optimizer algorithm to determine the safest convergence step.
\end{enumerate}

\begin{figure}[htbp]
\centering
\resizebox{0.9\linewidth}{!}{%
\begin{tikzpicture}[node distance=1.2cm, auto,
  box/.style={rectangle, draw, align=center, minimum width=7cm, minimum height=1cm},
  arrow/.style={-{Stealth}, thick}]

  \node (input) {\textbf{Input:} Experimental Probabilities};
  \node[box, below=of input] (init) {Initialize variable array $x_0$ \\ (Populations + Coherences) \\ and Evaluate Trace and Eigenvalues};
  \node[box, below=of init] (construct) {Construct Hermitian Matrix \\ $\hat{\rho}_{\text{guess}}$ from $x$ \\ and Evaluate Trace and Eigenvalues};
  \node[box, below=of construct] (slsqp) {SLSQP Engine: \\ Build local quadratic model, \\ check constraints, \\ compute safe step};
  \node[box, below=of slsqp] (eval) {Evaluate Likelihood Score};
  \node[box, below=of eval] (converge) {Has $\Delta x$ converged?};
  \node[box, below=of converge] (enforce) {Enforce valid quantum state \\ (Clean microscopic numeric artifacts)};
  \node[below=of enforce] (output) {\textbf{Output:} Final Valid Density Matrix $\hat{\rho}_{\text{final}}$};

  \draw[arrow] (input) -- (init);
  \draw[arrow] (init) -- (construct);
  \draw[arrow] (construct) -- (slsqp);
  \draw[arrow] (slsqp) -- (eval);
  \draw[arrow] (eval) -- (converge);
  \draw[arrow] (converge) -- node[right] {Yes} (enforce);
  \draw[arrow] (enforce) -- (output);
  \draw[arrow] (converge.west) -- node[above] {No (Update $x$)} ++(-2.5,0) |- (construct.west);
\end{tikzpicture}%
}
\caption{\justifying Flowchart of the Direct Hermitian parametrization procedure using the SLSQP optimizer, detailing initialization from experimental probabilities, iterative matrix construction and constraint checking, convergence evaluation, and final physical state enforcement.}
\label{fig:slsqp_flowchart}
\end{figure}
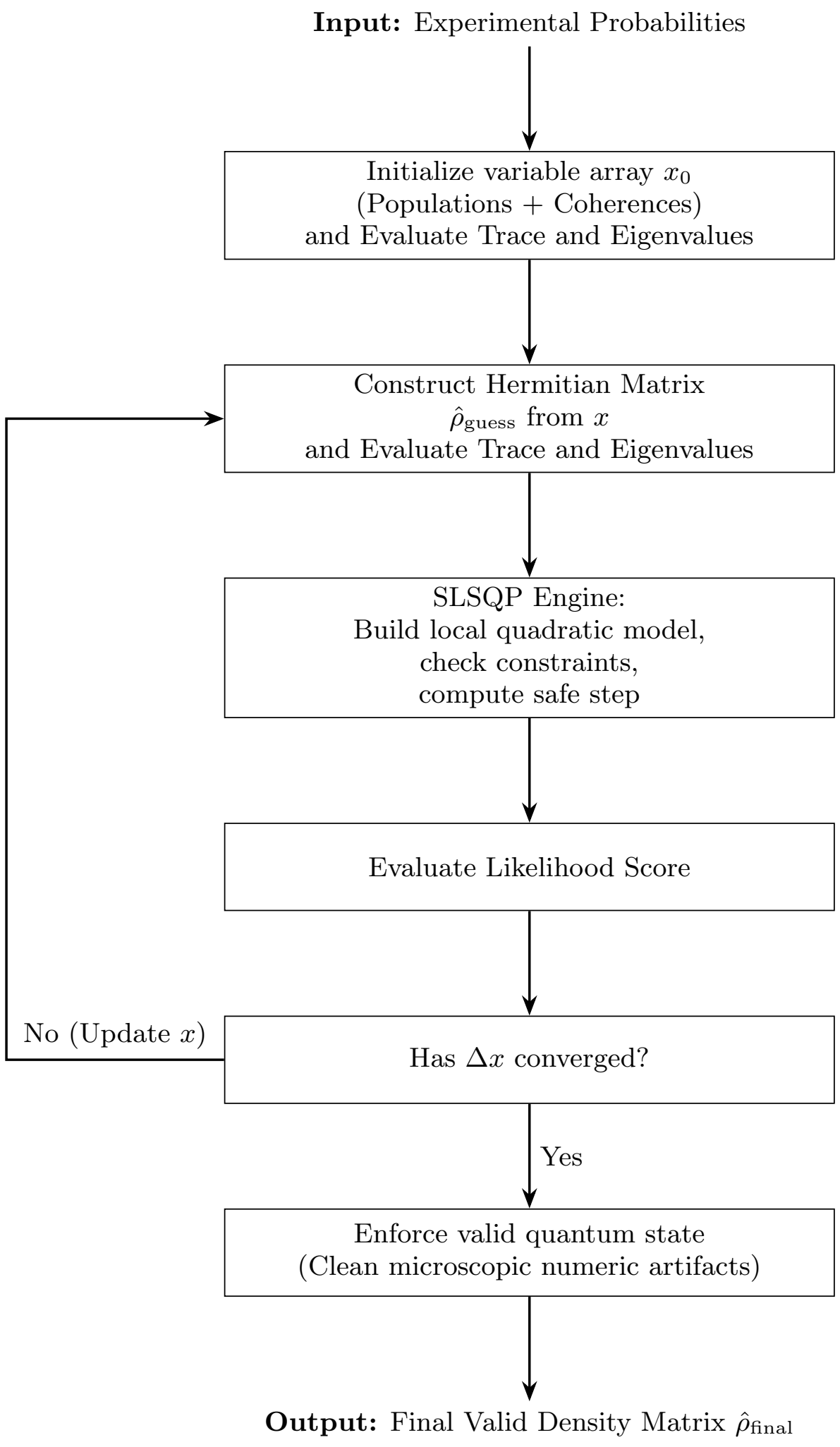

\subsection{Direct Hermitian Parametrization with SLSQP}
An alternative estimation approach utilizes Direct Hermitian parametrization coupled with a Sequential Least-Squares Programming (SLSQP) optimizer. The step-by-step workflow of this procedure is shown in Fig.~\ref{fig:slsqp_flowchart}.

As outlined in Fig.~\ref{fig:slsqp_flowchart}, the process begins with the Direct Hermitian Parametrization, where a one-dimensional array is constructed directly from the population terms and the specific upper triangular coherences that exceed the established threshold. For example, if evaluating a two-qubit system containing four population terms and a single relevant coherence term ($\rho_{12}$), the length of the 1-D array will be composed of the 4 population terms plus 2 variables representing the real and imaginary parts of the complex coherence term. This yields a total of 6 independent parameters that the optimizer must tune. The intermediate density matrix $\hat{\rho}_{\text{guess}}$ is then easily constructed from this 1-D array: the leading variables populate the diagonal, the appropriate variables fill the targeted upper triangular coherences, the lower triangular elements are obtained by conjugating the upper ones, and all other non-selected coherence entries are explicitly set to zero.

For initialization, the 1-D matrix is seeded with the initial experimental population probabilities extracted directly from $\hat{\rho}_{\text{raw}}$, while the coherence entries in the 1-D matrix begin at zero. 

The parameter tuning is subsequently driven by the SLSQP algorithm. SLSQP is a gradient-based numerical solver uniquely designed for mathematical problems requiring strict boundaries. For more details about the SLSQP optimizer algorithm, see Ref.~\cite{kraft1988}, which is also utilized by the IBM Quantum platform for such constrained numerical tasks. Because the constructed 1-D variable array consists of mathematically independent elements, an unrestricted optimizer could easily wander into unphysical state space. To prevent this, the SLSQP engine is bounded by rigid physical constraints: an equality constraint to ensure the Trace remains exactly 1 ($\text{Tr}(\hat{\rho}_{\text{guess}}) = 1$), and an inequality constraint verifying that the minimum eigenvalue of $\hat{\rho}_{\text{guess}}$ remains $\ge 0$ to satisfy positive semi-definiteness. 

Finally, the optimization is evaluated via a probability-weighted objective function closely related to the Gaussian likelihood approximation commonly utilized in standard Maximum Likelihood Estimation (MLE). For each experimentally measured projector $\hat{P}_k$, the probability predicted by a trial density matrix $\hat{\rho}$ is given by
\begin{equation}
p_k^{\mathrm{expected}}
=
\operatorname{Tr}(\hat{\rho}\hat{P}_k),
\label{eq:predicted_probability}
\end{equation}
whereas $p_k^{\mathrm{observed}}$ denotes the corresponding experimentally observed probability.

The objective function is defined as
\begin{equation}
\mathcal{L}(\hat{\rho})
=
\frac{1}{2}
\sum_{k\in\mathcal{M}}
\frac{
\left(
p_k^{\mathrm{expected}}-
p_k^{\mathrm{observed}}
\right)^2
}{
p_k^{\mathrm{expected}}
},
\label{eq:slsqp_objective}
\end{equation}
where $\mathcal{M}$ denotes the set of projective
measurements selected by the tQST protocol.

The comparison between the maximum-likelihood estimation process and the SLSQP-based estimation process is given in Table~\ref{tab:MLE-SLSQP}.

\begin{table*}[t]
\centering
\caption{Comparison of standard MLE-based and SLSQP-based reconstruction procedures.}
\label{tab:MLE-SLSQP}

\begingroup
\small
\renewcommand{\arraystretch}{1.2}
\setlength{\tabcolsep}{4pt}

\begin{tabular}{p{0.23\textwidth}p{0.34\textwidth}p{0.34\textwidth}}
\hline\hline

\textbf{Feature} &
\textbf{Standard MLE-based procedure} &
\textbf{SLSQP-based procedure} \\
\hline

Density matrix parametrization &
General triangular parametrization &
Threshold-restricted Hermitian parametrization \\ 
\hline

Optimization algorithm &
L-BFGS-B &
SLSQP \\ 
\hline

Objective function &
Gaussian-likelihood/Pearson-type weighted residual &
Pearson-type weighted least-squares objective \\
\hline

Allowed coherences &
All off-diagonal coherences are allowed to vary &
Only threshold-selected coherences are allowed to vary \\
\hline

Excluded coherences &
No off-diagonal coherences explicitly excluded &
Non-selected coherences explicitly set to zero \\
\hline

Positivity constraint &
Enforced through triangular parametrization &
Enforced through a minimum-eigenvalue inequality constraint \\
\hline
Trace constraint &
Enforced through normalization of $\hat{T}^\dagger \hat{T}$ &
Imposed as an equality constraint \\
\hline
Number of physical parameters (two qubits) &
$15$ &
$3+2k$, where $k$ is the number of retained
independent complex coherence pairs \\
\hline
Initial optimization values &
Random initialization using the original tQST implementation &
Measured diagonal populations with zero initial coherences \\
\hline
Prior knowledge of Bell-state identity &
Not required &
Not required \\

\hline\hline
\end{tabular}
\endgroup
\end{table*}

\bibliography{aapmsamp}% Produces the bibliography via BibTeX.

@article{Caruccio2025,
  author    = {Caruccio, Eugenio and Maragnano, Diego and Rodari, Giovanni and Picus, Davide and Garberoglio, Giovanni and Binosi, Daniele and Albiero, Riccardo and Di Giano, Niki and Ceccarelli, Francesco and Corrielli, Giacomo and Spagnolo, Nicol{\`o} and Osellame, Roberto and Dapor, Maurizio and Liscidini, Marco and Sciarrino, Fabio},
  title     = {Experimental verification of threshold quantum state tomography on a fully-reconfigurable photonic integrated circuit},
  journal   = {npj Quantum Information},
  volume    = {11},
  number    = {1},
  pages     = {173},
  year      = {2025},
  month     = nov,
  issn      = {2056-6387},
  doi       = {10.1038/s41534-025-01111-z},
  url       = {https://doi.org/10.1038/s41534-025-01111-z}
}

@article{10.1063/5.0219143,
    author = {Binosi, D. and Garberoglio, G. and Maragnano, D. and Dapor, M. and Liscidini, M.},
    title = {A tailor-made quantum state tomography approach},
    journal = {APL Quantum},
    volume = {1},
    number = {3},
    pages = {036112},
    year = {2024},
    month = {07},
    issn = {2835-0103},
    doi = {10.1063/5.0219143},
    url = {https://doi.org/10.1063/5.0219143},
}

@article{PhysRevLett.75.4337,
  title = {New High-Intensity Source of Polarization-Entangled Photon Pairs},
  author = {Kwiat, Paul G. and Mattle, Klaus and Weinfurter, Harald and Zeilinger, Anton and Sergienko, Alexander V. and Shih, Yanhua},
  journal = {Phys. Rev. Lett.},
  volume = {75},
  issue = {24},
  pages = {4337--4341},
  numpages = {0},
  year = {1995},
  month = {Dec},
  publisher = {American Physical Society},
  doi = {10.1103/PhysRevLett.75.4337},
  url = {https://link.aps.org/doi/10.1103/PhysRevLett.75.4337}
}

@article{PhysRevA.60.R773,
  title = {Ultrabright source of polarization-entangled photons},
  author = {Kwiat, Paul G. and Waks, Edo and White, Andrew G. and Appelbaum, Ian and Eberhard, Philippe H.},
  journal = {Phys. Rev. A},
  volume = {60},
  issue = {2},
  pages = {R773(R)--R776(R)},
  numpages = {0},
  year = {1999},
  month = {Aug},
  publisher = {American Physical Society},
  doi = {10.1103/PhysRevA.60.R773},
  url = {https://link.aps.org/doi/10.1103/PhysRevA.60.R773}
}

@article{PhysRevA.64.052312,
  title = {Measurement of qubits},
  author = {James, Daniel F. V. and Kwiat, Paul G. and Munro, William J. and White, Andrew G.},
  journal = {Phys. Rev. A},
  volume = {64},
  issue = {5},
  pages = {052312},
  numpages = {15},
  year = {2001},
  month = {Oct},
  publisher = {American Physical Society},
  doi = {10.1103/PhysRevA.64.052312},
  url = {https://link.aps.org/doi/10.1103/PhysRevA.64.052312}
}

@article{PhysRevLett.105.150401,
  title = {Quantum State Tomography via Compressed Sensing},
  author = {Gross, David and Liu, Yi-Kai and Flammia, Steven T. and Becker, Stephen and Eisert, Jens},
  journal = {Phys. Rev. Lett.},
  volume = {105},
  issue = {15},
  pages = {150401},
  numpages = {4},
  year = {2010},
  month = {Oct},
  publisher = {American Physical Society},
  doi = {10.1103/PhysRevLett.105.150401},
  url = {https://link.aps.org/doi/10.1103/PhysRevLett.105.150401}
}

@article{Qi2017,
  author    = {Qi, Bo and Hou, Zhibo and Wang, Yuanlong and Dong, Daoyi and Zhong, Han-Sen and Li, Li and Xiang, Guo-Yong and Wiseman, Howard M. and Li, Chuan-Feng and Guo, Guang-Can},
  title     = {Adaptive quantum state tomography via linear regression estimation: Theory and two-qubit experiment},
  journal   = {npj Quantum Information},
  volume    = {3},
  number    = {1},
  pages     = {19},
  year      = {2017},
  month     = apr,
  issn      = {2056-6387},
  doi       = {10.1038/s41534-017-0016-4},
  url       = {https://doi.org/10.1038/s41534-017-0016-4}
}

@article{PhysRevA.111.032436,
  title = {Enhanced compressive threshold quantum state tomography for qudit systems},
  author = {Garberoglio, Giovanni and Dapor, Maurizio and Maragnano, Diego and Liscidini, Marco and Binosi, Daniele},
  journal = {Phys. Rev. A},
  volume = {111},
  issue = {3},
  pages = {032436},
  numpages = {13},
  year = {2025},
  month = {Mar},
  publisher = {American Physical Society},
  doi = {10.1103/PhysRevA.111.032436},
  url = {https://link.aps.org/doi/10.1103/PhysRevA.111.032436}
}

@article{Sergienko01072003,
author = {Sergienko, Av and Jaeger, Gs},
title = {Quantum information processing and precise optical measurement with entangled-photon pairs},
journal = {Contemporary Physics},
volume = {44},
number = {4},
pages = {341--356},
year = {2003},
publisher = {Taylor \& Francis},
doi = {10.1080/0010751031000102711},
URL = {https://doi.org/10.1080/0010751031000102711}
}

@article{Bourennane2003,
  author    = {Bourennane, Mohamed and Eibl, Manfred and Gaertner, Sascha and Kiesel, Nikolai and Kurtsiefer, Christian and {\.Z}ukowski, Marek and Weinfurter, Harald},
  title     = {Multiphoton entanglement and interferometry},
  journal   = {Fortschritte der Physik},
  volume    = {51},
  number    = {4--5},
  pages     = {273--279},
  year      = {2003},
  publisher = {Wiley-VCH},
  doi       = {10.1002/prop.200310039}
}

@article{y1pg-nx49,
  title = {Accurate and precise optical phase sensor based on a nonlinear quantum Sagnac interferometer},
  author = {Dalidet, Romain and Labont{\'e}, Laurent and Sauder, Gregory and Tanzilli, S{\'e}bastien and Martin, Anthony},
  journal = {Phys. Rev. Res.},
  volume = {7},
  issue = {3},
  pages = {033175},
  numpages = {7},
  year = {2025},
  month = {Aug},
  publisher = {American Physical Society},
  doi = {10.1103/y1pg-nx49},
  url = {https://link.aps.org/doi/10.1103/y1pg-nx49}
}

@article{Thik_2026,
doi = {10.1088/2040-8986/aea6f0},
url = {https://doi.org/10.1088/2040-8986/aea6f0},
year = {2026},
month = {sep},
publisher = {IOP Publishing},
volume = {28},
number = {9},
pages = {095201},
author = {Thik, Gayatri and Loyal, Amit and K, Srinivasan and G, Raghavan},
title = {On-demand generation of all four bell states using a single PPKTP entangled photon source},
journal = {Journal of Optics}
}

@Inbook{Altepeter2004,
author="Altepeter, Joseph B.
and James, Daniel F.V.
and Kwiat, Paul G.",
editor="Paris, Matteo
and {\v{R}}eh{\'a}{\v{c}}ek, Jaroslav",
title="4 Qubit Quantum State Tomography",
bookTitle="Quantum State Estimation",
year="2004",
publisher="Springer Berlin Heidelberg",
address="Berlin, Heidelberg",
pages="113--145",
isbn="978-3-540-44481-7",
doi="10.1007/978-3-540-44481-7_4",
url="https://doi.org/10.1007/978-3-540-44481-7_4"
}

@article{Sundar:25,
author = {S. Sundar and M. V. Jabir and D. Razansky},
journal = {J. Opt. Soc. Am. B},
number = {8},
pages = {1764--1770},
publisher = {Optica Publishing Group},
title = {Variable Bell states in a polarization-entangled photon source},
volume = {42},
month = {Aug},
year = {2025},
url = {https://opg.optica.org/josab/abstract.cfm?URI=josab-42-8-1764},
doi = {10.1364/JOSAB.560031},
}

@inproceedings{10.1117/12.504402,
author = {Paul G. Kwiat and Joseph Altepeter and Julio Barreiro and David A. Branning and Evan R. Jeffrey and Nicholas Peters and Aaron P. VanDevender},
title = {{Optical technologies for quantum information science}},
volume = {5161},
booktitle = {Quantum Communications and Quantum Imaging},
editor = {Ronald E. Meyers and Yanhua Shih},
organization = {International Society for Optics and Photonics},
publisher = {SPIE},
pages = {87 -- 100},
year = {2004},
doi = {10.1117/12.504402},
URL = {https://doi.org/10.1117/12.504402}
}

@article{Lu2022Bayesian,
  author    = {Lu, Hsuan-Hao and Myilswamy, Karthik V. and Bennink, Ryan S. and Seshadri, Suparna and Alshaykh, Mohammed S. and Liu, Junqiu and Kippenberg, Tobias J. and Leaird, Daniel E. and Weiner, Andrew M. and Lukens, Joseph M.},
  title     = {Bayesian tomography of high-dimensional on-chip biphoton frequency combs with randomized measurements},
  journal   = {Nature Communications},
  volume    = {13},
  number    = {1},
  pages     = {4338},
  year      = {2022},
  publisher = {Nature Publishing Group},
  doi       = {10.1038/s41467-022-31639-z}
}

@article{BUZEK1998454,
title = {Reconstruction of Quantum States of Spin Systems: From Quantum Bayesian Inference to Quantum Tomography},
journal = {Annals of Physics},
volume = {266},
number = {2},
pages = {454-496},
year = {1998},
issn = {0003-4916},
doi = {https://doi.org/10.1006/aphy.1998.5802},
url = {https://www.sciencedirect.com/science/article/pii/S000349169895802X},
author = {V. Bužek and R. Derka and G. Adam and P.L. Knight}
}

@Inbook{Greenberger1989,
author="Greenberger, Daniel M.
and Horne, Michael A.
and Zeilinger, Anton",
editor="Kafatos, Menas",
title="Going Beyond Bell's Theorem",
bookTitle="Bell's Theorem, Quantum Theory and Conceptions of the Universe",
year="1989",
publisher="Springer Netherlands",
address="Dordrecht",
pages="69--72",
isbn="978-94-017-0849-4",
doi="10.1007/978-94-017-0849-4_10",
url="https://doi.org/10.1007/978-94-017-0849-4_10"
}

@article{PhysRevA.62.062314,
  title = {Three qubits can be entangled in two inequivalent ways},
  author = {D{\"u}r, W. and Vidal, G. and Cirac, J. I.},
  journal = {Phys. Rev. A},
  volume = {62},
  issue = {6},
  pages = {062314},
  numpages = {12},
  year = {2000},
  month = {Nov},
  publisher = {American Physical Society},
  doi = {10.1103/PhysRevA.62.062314},
  url = {https://link.aps.org/doi/10.1103/PhysRevA.62.062314}
}

@article{10.1063/5.0023103,
    author = {Anwar, Ali and Perumangatt, Chithrabhanu and Steinlechner, Fabian and Jennewein, Thomas and Ling, Alexander},
    title = {Entangled photon-pair sources based on three-wave mixing in bulk crystals},
    journal = {Review of Scientific Instruments},
    volume = {92},
    number = {4},
    pages = {041101},
    year = {2021},
    month = {04},
    issn = {0034-6748},
    doi = {10.1063/5.0023103},
    url = {https://doi.org/10.1063/5.0023103},
}

@article{PhysRevLett.90.200403,
  title = {Experimental Observation of Four-Photon Entanglement from Parametric Down-Conversion},
  author = {Eibl, Manfred and Gaertner, Sascha and Bourennane, Mohamed and Kurtsiefer, Christian and {\ifmmode \dot{Z}\else \.{Z}\fi{}}ukowski, Marek and Weinfurter, Harald},
  journal = {Phys. Rev. Lett.},
  volume = {90},
  issue = {20},
  pages = {200403},
  numpages = {4},
  year = {2003},
  month = {May},
  publisher = {American Physical Society},
  doi = {10.1103/PhysRevLett.90.200403},
  url = {https://link.aps.org/doi/10.1103/PhysRevLett.90.200403}
}

@article{PhysRevA.64.010102,
  title = {Four-photon entanglement from down-conversion},
  author = {Weinfurter, Harald and {\ifmmode \dot{Z}\else \.{Z}\fi{}}ukowski, Marek},
  journal = {Phys. Rev. A},
  volume = {64},
  issue = {1},
  pages = {010102(R)},
  numpages = {4},
  year = {2001},
  month = {Jun},
  publisher = {American Physical Society},
  doi = {10.1103/PhysRevA.64.010102},
  url = {https://link.aps.org/doi/10.1103/PhysRevA.64.010102}
}

@article{Gaertner2003High,
  author  = {Gaertner, S. and Bourennane, M. and Eibl, M. and Kurtsiefer, C. and Weinfurter, H.},
  title   = {High-fidelity source of four-photon entanglement},
  journal = {Applied Physics B},
  volume  = {77},
  number  = {8},
  pages   = {803--807},
  year    = {2003},
  doi     = {10.1007/s00340-003-1321-5}
}

@book{nielsen2010quantum,
  author    = {Nielsen, Michael A. and Chuang, Isaac L.},
  title     = {Quantum Computation and Quantum Information: 10th Anniversary Edition},
  publisher = {Cambridge University Press},
  year      = {2010},
  isbn      = {9781139495486},
  doi       = {10.1017/CBO9780511976667}
}

@article{PhysRevA.68.012305,
  title = {Quantum inference of states and processes},
  author = {Je{\ifmmode \check{z}\else \v{z}\fi{}}ek, Miroslav and Fiur{\'a}{\ifmmode \check{s}\else \v{s}\fi{}}ek, Jarom{\'{\i}}r and Hradil, Zden{\ifmmode \check{e}\else \v{e}\fi{}}k},
  journal = {Phys. Rev. A},
  volume = {68},
  issue = {1},
  pages = {012305},
  numpages = {7},
  year = {2003},
  month = {Jul},
  publisher = {American Physical Society},
  doi = {10.1103/PhysRevA.68.012305},
  url = {https://link.aps.org/doi/10.1103/PhysRevA.68.012305}
}

@article{PhysRevLett.94.070402,
  title = {Full Characterization of a Three-Photon Greenberger-Horne-Zeilinger State Using Quantum State Tomography},
  author = {Resch, K. J. and Walther, P. and Zeilinger, A.},
  journal = {Phys. Rev. Lett.},
  volume = {94},
  issue = {7},
  pages = {070402},
  numpages = {4},
  year = {2005},
  month = {Feb},
  publisher = {American Physical Society},
  doi = {10.1103/PhysRevLett.94.070402},
  url = {https://link.aps.org/doi/10.1103/PhysRevLett.94.070402}
}

@article{PhysRevA.61.010304,
  title = {Maximum-likelihood estimation of the density matrix},
  author = {Banaszek, K. and D'Ariano, G. M. and Paris, M. G. A. and Sacchi, M. F.},
  journal = {Phys. Rev. A},
  volume = {61},
  issue = {1},
  pages = {010304(R)},
  numpages = {4},
  year = {1999},
  month = {Dec},
  publisher = {American Physical Society},
  doi = {10.1103/PhysRevA.61.010304},
  url = {https://link.aps.org/doi/10.1103/PhysRevA.61.010304}
}

@ARTICLE{5238742,
  author={Hurley, Niall and Rickard, Scott},
  journal={IEEE Transactions on Information Theory}, 
  title={Comparing Measures of Sparsity}, 
  year={2009},
  volume={55},
  number={10},
  pages={4723-4741},
  doi={10.1109/TIT.2009.2027527}}

@book{HornJohnson2012,
  author    = {Horn, Roger A. and Johnson, Charles R.},
  title     = {Matrix Analysis},
  edition   = {2},
  publisher = {Cambridge University Press},
  address   = {Cambridge},
  year      = {2012},
  isbn      = {978-0-521-83940-2},
  doi       = {10.1017/CBO9781139020411},
  url       = {https://doi.org}
}

@article{BERA2025132401,
title = {Bright source of degenerate polarization-entangled photons using type-0 PPKTP crystal: Effects of accidental coincidences},
journal = {Optics Communications},
volume = {596},
pages = {132401},
year = {2025},
issn = {0030-4018},
doi = {https://doi.org/10.1016/j.optcom.2025.132401},
url = {https://www.sciencedirect.com/science/article/pii/S0030401825009290},
author = {Kousik Bera and Pawan Kumar Dubey and Ankush Kumar and Manoj Jha}
}

@article{Terashima2018,
  author  = {Terashima, Haruka and Kobayashi, Satoshi and Tsubakiyama, Takaho and Sanaka, Kaoru},
  title   = {Quantum interferometric generation of polarization entangled photons},
  journal = {Scientific Reports},
  year    = {2018},
  volume  = {8},
  number  = {1},
  pages   = {15733},
  doi     = {10.1038/s41598-018-33876-z},
  url     = {https://doi.org/10.1038/s41598-018-33876-z},
  issn    = {2045-2322}
}

@article{article,
author = {Cai, Neng and Cai, Wu-Hao and Wang, Shun and Li, Fang and Shimizu, Ryosuke and Jin, Ruibo},
year = {2021},
month = {12},
pages = {77-82},
title = {Broadband-laser-diode pumped periodically poled potassium titanyl phosphate-Sagnac polarization-entangled photon source},
volume = {39},
journal = {Journal of the Optical Society of America B},
doi = {10.1364/JOSAB.437808}
}

@article{Li:15,
author = {Yan Li and Zhi-Yuan Zhou and Dong-Sheng Ding and Bao-Sen Shi},
journal = {Opt. Express},
number = {22},
pages = {28792--28800},
publisher = {Optica Publishing Group},
title = {CW-pumped telecom band polarization entangled photon pair generation in a Sagnac interferometer},
volume = {23},
month = {Nov},
year = {2015},
url = {https://opg.optica.org/oe/abstract.cfm?URI=oe-23-22-28792},
doi = {10.1364/OE.23.028792},
}

@article{Lee2015,
  author  = {Lee, Hee Jung and Kim, Heonoh and Lee, Sang Min and Moon, Han Seb},
  title   = {Indistinguishability of photon pair in a periodically poled {KTiOPO$_4$}},
  journal = {Applied Physics B},
  year    = {2015},
  volume  = {121},
  number  = {4},
  pages   = {541--547},
  doi     = {10.1007/s00340-015-6270-2},
  url     = {https://doi.org},
  issn    = {1432-0649}
}

@article{PhysRevLett.70.1895,
  title = {Teleporting an unknown quantum state via dual classical and Einstein-Podolsky-Rosen channels},
  author = {Bennett, Charles H. and Brassard, Gilles and Cr{\'e}peau, Claude and Jozsa, Richard and Peres, Asher and Wootters, William K.},
  journal = {Phys. Rev. Lett.},
  volume = {70},
  issue = {13},
  pages = {1895--1899},
  numpages = {0},
  year = {1993},
  month = {Mar},
  publisher = {American Physical Society},
  doi = {10.1103/PhysRevLett.70.1895},
  url = {https://link.aps.org/doi/10.1103/PhysRevLett.70.1895}
}

@article{PhysRevLett.69.2881,
  title = {Communication via one- and two-particle operators on Einstein-Podolsky-Rosen states},
  author = {Bennett, Charles H. and Wiesner, Stephen J.},
  journal = {Phys. Rev. Lett.},
  volume = {69},
  issue = {20},
  pages = {2881--2884},
  numpages = {0},
  year = {1992},
  month = {Nov},
  publisher = {American Physical Society},
  doi = {10.1103/PhysRevLett.69.2881},
  url = {https://link.aps.org/doi/10.1103/PhysRevLett.69.2881}
}

@article{PhysRevLett.67.661,
  title = {Quantum cryptography based on Bell's theorem},
  author = {Ekert, Artur K.},
  journal = {Phys. Rev. Lett.},
  volume = {67},
  issue = {6},
  pages = {661--663},
  numpages = {0},
  year = {1991},
  month = {Aug},
  publisher = {American Physical Society},
  doi = {10.1103/PhysRevLett.67.661},
  url = {https://link.aps.org/doi/10.1103/PhysRevLett.67.661}
}

@article{10.1116/5.0007577,
    author = {Polino, Emanuele and Valeri, Mauro and Spagnolo, Nicolò and Sciarrino, Fabio},
    title = {Photonic quantum metrology},
    journal = {AVS Quantum Science},
    volume = {2},
    number = {2},
    pages = {024703},
    year = {2020},
    month = {06},
    issn = {2639-0213},
    doi = {10.1116/5.0007577},
    url = {https://doi.org/10.1116/5.0007577},
    
}

@article{Byrd1995LBFGSB,
  author  = {Byrd, Richard H. and Lu, Peihuang and Nocedal, Jorge and Zhu, Ciyou},
  title   = {A Limited Memory Algorithm for Bound Constrained Optimization},
  journal = {SIAM Journal on Scientific Computing},
  year    = {1995},
  volume  = {16},
  number  = {5},
  pages   = {1190--1208},
  doi     = {10.1137/0916069}
}

@article{Wootters1998,
  author  = {Wootters, William K.},
  title   = {Entanglement of Formation of an Arbitrary State of Two Qubits},
  journal = {Physical Review Letters},
  year    = {1998},
  volume  = {80},
  number  = {10},
  pages   = {2245--2248},
  doi     = {10.1103/PhysRevLett.80.2245}
}

@book{kraft1988,
  title={A Software Package for Sequential Quadratic Programming},
  author={Kraft, D.},
  series={Deutsche Forschungs- und Versuchsanstalt f{\"u}r Luft- und Raumfahrt K{\"o}ln: Forschungsbericht},
  url={https://books.google.co.in/books?id=4rKaGwAACAAJ},
  year={1988},
  publisher={Wiss. Berichtswesen d. DFVLR}
}

\end{document}